\RequirePackage{snapshot}

\documentclass[a4paper,fleqn,usenatbib]{mnras}

\usepackage{newtxtext,newtxmath}

\usepackage[T1]{fontenc}
\usepackage{ae,aecompl}

\usepackage{graphicx}	
\usepackage{amsmath}	
\usepackage{amssymb}	
\usepackage{upgreek}
\usepackage{xspace}
\usepackage{ulem}
\usepackage{relsize}
\usepackage{tikz}
\usetikzlibrary{positioning}
\usetikzlibrary{calc}
\usetikzlibrary{intersections}
\hypersetup{draft}

\newcommand{\ag}{{\it AGILE}\xspace}
\newcommand{\fer}{{\it Fermi} LAT\xspace}
\newcommand{\cyg}{Cygnus X-3\xspace}
\newcommand{\cygl}{Cygnus X-1\xspace}

\newcommand{\muq}{\(\upmu\)Q\xspace}
\newcommand{\be}{\begin{equation}}
\newcommand{\ee}{\end{equation}}
\newcommand{\ba}{\begin{eqnarray}}
\newcommand{\ea}{\end{eqnarray}}

\newcommand{\ergs}{\ensuremath{\mathrm{erg\,s^{-1}}}}

\newcommand{\bp}{\ensuremath{\mathbf{p}}}
\newcommand{\br}{\ensuremath{\mathbf{r}}}
\newcommand{\Db}{\ensuremath{{\cal D}}}
\newcommand{\ti}[1]{\ensuremath{\tilde #1}}
\newcommand{\ve}{\ensuremath{\varepsilon}}

\newcommand{\rikkyo}{{Department of Physics, Rikkyo University, Nishi-Ikebukuro 3-34-1, Toshima-ku, Tokyo 171-8501, Japan}}
\newcommand{\riken}{{RIKEN iTHEMS, Hirosawa 2-1, Wako, Saitama 351-0198, Japan}}

\newcommand{\UB}{Departament de F\'{i}sica Qu\`antica i Astrof\'{i}sica, Institut de Ci\`encies del Cosmos (ICC), Universitat de Barcelona (IEEC-UB), Mart\'{i} i Franqu\`es 1, E08028 Barcelona, Spain}

\title[IC from Relativistic Jets in Binary Systems]{Inverse Compton Emission from Relativistic Jets in Binary Systems}

\author[D. Khangulyan et al.]{
Dmitry Khangulyan$^{1,2}$\thanks{E-mail: d.khangulyan@rikkyo.ac.jp},
Valent\'i Bosch-Ramon$^{3}$
and Yasunobu Uchiyama$^{1}$\\
$^{1}$\rikkyo\\
$^{2}$\riken\\
$^{3}$\UB
}

\date{Accepted XXX. Received YYY; in original form ZZZ}

\pubyear{2018}

\begin{document}
\label{firstpage}
\pagerange{\pageref{firstpage}--\pageref{lastpage}}
\maketitle

\begin{abstract}
The gamma-ray emission detected from several microquasars can be produced by relativistic electrons emitting through inverse Compton scattering. In particular, the GeV emission detected from \cyg, and its orbital phase dependence, strongly suggest that the emitting electrons are accelerated in a relativistic jet, and that the optical companion provides the dominant target. Here, we study the effects related to particle transport in the framework of the relativistic jet scenario. We find that even in the most compact binary systems, with parameters similar to \cyg,  particle transport can have a substantial influence on the GeV lightcurve unless the jet is slow, \(\beta < 0.7\). In more extended binary systems, strong impact of particle transport is nearly unavoidable. 
Thus,
even for a very compact system such as \cyg, particle transport significantly affects the ability of one-zone models to infer the properties of the gamma-ray production site based on the shape on the GeV lightcurve. We conclude that a detailed study of the gamma-ray spectrum can further constrain the structure and other properties of the gamma-ray emitter in \cyg, although such a study should account for gamma-gamma attenuation, since it may strongly affect the spectrum above \(5\rm\,GeV\).
\end{abstract}


\section{Introduction}

Microquasars (\muq) are binary systems that host a companion star and an accreting compact object (CO) from which jets are produced. Several microquasars have been detected in the GeV gamma-ray range with \ag\, and \fer\, \citep{2009Natur.462..620T,2009Sci...326.1512F,2011ApJ...733L..20W,2012A&A...545A.110P,2012A&A...538A..63B,2012MNRAS.421.2947C,2013MNRAS.434.2380M,2013ApJ...775...98B,2016A&A...596A..55Z,2017ApJ...839...84P}. The variability found in the GeV emission in some of these sources is consistent with inverse Compton (IC) scattering of stellar photons by relativistic electrons accelerated in the jets \citep[e.g.,][]{2010MNRAS.404L..55D,2016A&A...596A..55Z,2018MNRAS.479.4399Z}. The IC origin of the gamma-ray emission detected from \muq is supported by arguments based on higher efficiency of leptonic radiation mechanisms, as compared to hadronic ones, under conditions of compact binary systems \citep{2009IJMPD..18..347B}. 

If the dominant target photon field is provided by the stellar companion, IC scattering will be strongly anisotropic \citep[see, e.g.,][]{2005AIPC..745..359K,2008MNRAS.383..467K}, and the scattering angle will change along the orbit. This variability of the scattering angle is imprinted in the emission intensity, and may be the dominant factor shaping the GeV lightcurve \citep[e.g.,][for \cyg]{2010MNRAS.404L..55D}. The specific dependence of the scattering angle on the orbital phase is determined by the jet and counter-jet orientations, and the location of the acceleration and the emission sites in the jet. Thus, gamma-ray light curves can help in constraining the emitter location in \muq.

\cyg is the brightest and best studied gamma-ray emitting \muq \citep[e.g.][]{2009Natur.462..620T,2009Sci...326.1512F}.
The high luminosity of this source may favour, from energetic arguments, relativistic jet velocities, as they could alleviate the demanding energy requirements through Doppler boosting. In such a jet, the non-thermal distribution of particles and their emission would be significantly affected by relativistic effects. Nevertheless, a highly relativistic jet is somewhat in tension with \fer\, data in the context of a one-zone IC emitter \citep{2010MNRAS.404L..55D,2018MNRAS.479.4399Z}. On the other hand, radio VLBI observations of the jets of \cyg, from milliarcsecond-to-arcsecond scales (\(\sim 10-10^4\)~AU), favour an at least moderately relativistic jet  \citep{2001ApJ...553..766M,2001A&A...375..476M}, which may point to an even more relativistic flow on the scales of the binary (\(\sim 0.1\)~AU).

In this paper, we derive the formulas for the IC emission from a relativistic jet using the distribution function of electrons in the phase space \((\br,\bp)\), with \(\br\) and \(\bp\) being the particle spatial and momentum coordinates in the laboratory reference frame (RF), respectively. This function is a Lorentz invariant, which allows us to avoid cumbersome RF transformations {in the case when the contribution from synchrotron self-Compton (SSC) is negligible\footnote{Note that in the case of a very clumpy jets, the SSC mechanism may provide a non-negligible contribution \citep[see][for the case of \cygl]{2017MNRAS.471.3657Z}}}. This approach also allows us to obtain the results in a form that consistently describes the advection and radiation of gamma rays by particles in the case of an extended emitter. In the derivation, we account both for the transformation of the particle distribution to the laboratory frame, and for the impact of relativistic effects on the particle cooling in the plasma frame. We obtain an analytic solution for the invariant distribution function under the assumption of dominant Thomson IC losses, and numerically compute the IC radiation accounting for changes in the target density and scattering angle along the jet. We discuss the impact of the synchrotron and adiabatic losses, and characterize the conditions when synchrotron losses dominate, under which an analytic solution for the particle distribution can be obtained.

An approach based on the invariant distribution function was earlier suggested to describe the beaming pattern of the external IC emission produced by blobs moving relativistically in blazar jets \citep{2001ApJ...561..111G}. This approach was later applied to study variable IC emission in binary systems \citep[see, e.g.,][]{2002A&A...385L..10K,2002A&A...388L..25G,2002A&A...393L..61R}. In contrast to these studies, in our paper we consider the emission produced in a jet which implies a different beaming pattern, as compared to an emitting blob. Another difference with the calculations presented by \citet{2001ApJ...561..111G} is that we use the invariant distribution function to describe the propagation and cooling of relativistic electrons in an extended emitter, which appears to be an important factor for interpreting the gamma-ray emission detected from gamma-ray binary systems.

As compared to other models, which involve extended emitters in \muq \citep[see, e.g.,][]{2010MNRAS.403.1457V,2012A&A...538A..97V,2014MNRAS.440.2238Z,2014MNRAS.442.3243Z,2015A&A...584A..95P} and rely on the conventional approach with RF transformations, our method significantly simplifies the computation of the external IC emission. Thus, this paper allows us to extend the existing models focusing on the GeV gamma-ray emission from \muq \citep[e.g.,][]{2010MNRAS.404L..55D,2012MNRAS.421.2956Z,2018MNRAS.479.4399Z}, and to study consistently the influence of particle advection on the gamma-ray spectra and lightcurves. 

Under conditions typical for \muq, in the TeV energy band the Klein-Nishina regime and gamma-gamma attenuation can affect the IC scattering and propagation of gamma rays, respectively \citep[see, e.g.,][]{2009IJMPD..18..347B}. In some systems with  particularly hot stellar companions, e.g. as \cyg, these effects may influence the production of GeV gamma rays \citep[see, e.g.,][]{1987ApJ...322..838P,1993MNRAS.260..681M,1997A&A...322..523B,2011A&A...529A.120C,2012MNRAS.421..512S}. Therefore, we also consider the influence of the Klein-Nishina effect on the electron transport and the impact of the gamma-gamma absorption on the spectrum adopting system parameters similar to \cyg.

\section{Upscattering of External Photons by a Relativistic Jet}

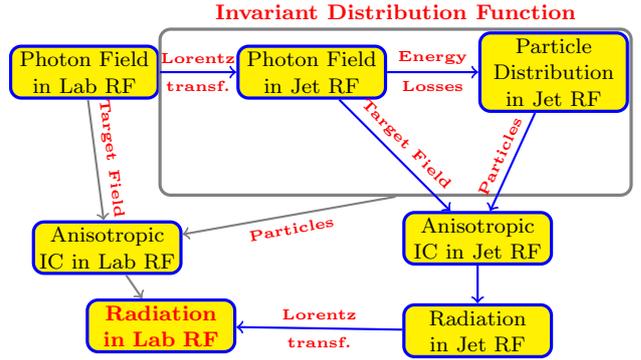
\begin{figure}
  \begin{tikzpicture}[
    blocks/.style={rectangle, rounded corners, draw=blue, fill=yellow, very thick, font=\relsize{0}, text width=1.8cm, minimum height=2em, inner sep=2pt, text centered}
    ]
    \def\xsize{3cm}
    \def\ysize{3cm}
    \node[blocks] (first)  at (0,0) {
Photon Field in Lab RF
};
\node[blocks, xshift=2cm] (second)  at (first.east) {
Photon Field in Jet RF
};
\node[blocks, right, xshift=1.2cm] (third)  at (second.east) {
Particle Distribution in Jet RF
};

\node[blocks, below, xshift=-1cm, yshift=-1.3cm] (ic1)  at (third.south) {
Anisotropic IC in Jet RF
};
\node[draw=gray,very thick, rounded corners, rectangle, minimum width=6.2cm, minimum height = 2.2cm, anchor = north east, below left, xshift=0.05cm, yshift=0.05cm] (invariant) at (third.north east){};
\node[anchor=south, above] at (invariant.north) {
  \color{red}\bf Invariant Distribution Function
  };
\node[blocks,draw=blue, below, xshift=0.3cm, yshift=-1.6cm] (ic2)  at (first.south) {
Anisotropic IC in Lab RF
};

\node[blocks, below, yshift=-0.5cm] (fourth)  at (ic1.south) {
Radiation in Jet RF
};
\node[blocks, left, xshift=2cm, yshift=-3cm] (fith)  at (first.south) {
{\bf \color{red} Radiation in Lab RF}
};
\path  (first)     	edge[->,blue,thick] node[anchor=south,above,font=\relsize{-2}]{\color{red}\bf Lorentz} node[anchor=south,below,font=\relsize{-2}]{\color{red}\bf transf.} (second);
\path  (second)     	edge[->,blue,thick] node[anchor=south,above,font=\relsize{-2}]{\color{red}\bf Energy} node[anchor=south,below,font=\relsize{-2}]{\color{red}\bf  Losses} (third);
\path  (third)     	edge[->,blue,thick] node[anchor=south,above,font=\relsize{-2},sloped]{\color{red}\bf Particles}  (ic1); 
\path  (second)     	edge[->,blue,thick] node[anchor=south,above,font=\relsize{-2},sloped]{\color{red}\bf Target Field}  (ic1);
\path (ic1) edge[->,blue,thick] (fourth);
\path  (fourth)     	edge[->,blue,thick] node[anchor=south,above,font=\relsize{-2}]{\color{red}\bf Lorentz} node[anchor=south,below,font=\relsize{-2}]{\color{red}\bf transf.} (fith);
\path  (invariant.south)  edge[->,gray,thick] node[anchor=north,below,font=\relsize{-2},sloped]{\color{red}\bf Particles}  (ic2); 
\path  (first)     	edge[->,gray,thick] node[anchor=south,above,font=\relsize{-2},sloped]{\color{red}\bf Target Field}  (ic2);
\path (ic2) edge[->,gray,thick] (fith);
\end{tikzpicture}
\caption{Algorithm for calculating the IC emission using the conventional approach and the Lorentz invariant distribution function.}
\label{fig:integral}
\end{figure}

Conventionally the IC radiation from relativistic sources is computed as follows. First one transforms the radiation field to the jet frame. Then, the distribution of high-energy electrons in that frame is obtained, from which the IC emission is computed. Finally, using the relativistic transformation of the radiation to the laboratory frame, one tranforms the emission to that frame \citep[for \muq see, e.g.,][]{2010A&A...516A..18D,2012MNRAS.421.2956Z}. Although this method is straightforward, in some contexts it may be more convenient to follow a different path: using Lorentz invariant quantities allows avoiding several frame transformations and provide the results in a form that illustrates the influence of different parameters clearly \cite[see, e.g.,][]{2001ApJ...561..111G}. The difference between these two approaches is sketched in Fig.~\ref{fig:integral}.

To describe the properties of non-thermal particles in the jet, we use the distribution function in the phase space:
\be
d N = f(t,\br,\bp)d^3\br d^3\bp\,.
\ee
We follow a general notation policy in which the uppercase ``N'' refers to {\itshape number} of particles, i.e., a dimensionless quantity, and the lowercase letters, e.g., ``n'', ``f'', to {\itshape densities} or {\itshape (differential) distributions}, i.e., \(dN = n dX\), where \(X\) is some quantity or a set of quantities. 

\begin{figure}
  \includegraphics[width=\columnwidth]{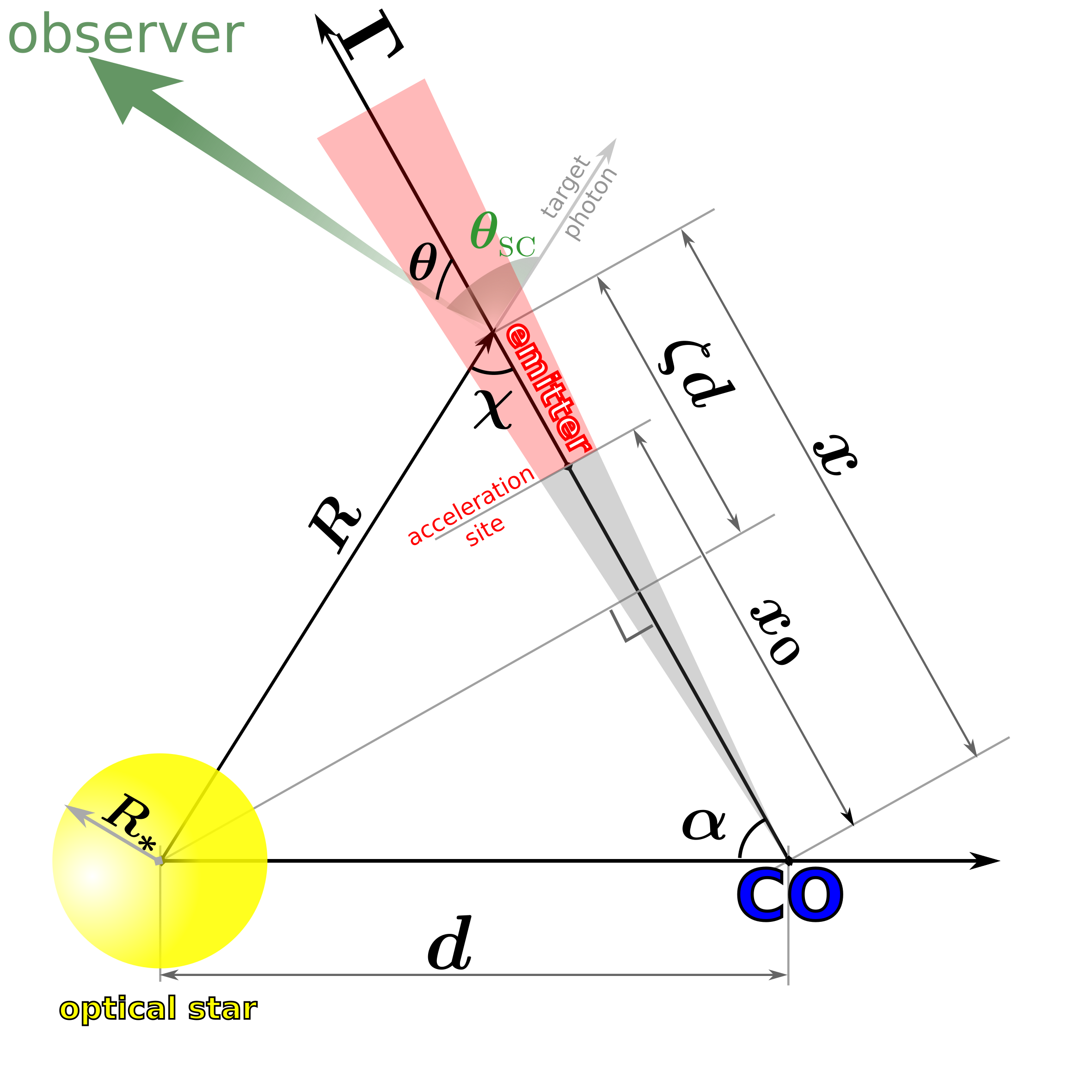}
  \caption{Sketch of the considered jet geometry.}
  \label{fig:geom}
\end{figure}

We assume that non-thermal particles are confined in a narrow jet. Thus, in a coordinate system in which the jet is directed along the \(x\)-axis, and the system origin is at the CO location, the distribution function should depend on the \(x\)-coordinate only, and on the particle momenta. The acceleration site is located at a distance \(x=x_0\) from the CO, and from there non-thermal particles are advected downstream along a relativistic jet that moves with bulk velocity \(\beta=V/c\) and Lorentz factor \(\Gamma=1/\sqrt{1-\beta^2}\) (see Fig.~\ref{fig:geom}, where the model geometry is illustrated).

The acceleration site can be associated for instance with a recollimation shock, and may vary with time: \(x_0=x_0(t)\) \citep[see, e.g.,][]{2010A&A...512L...4P}. However, for the sake of simplicity, it is typically assumed that this distance changes slowly, as compared to the characteristic advection/cooling times, which are also significantly shorter than the binary orbital period \citep[e.g.,][]{2009IJMPD..18..347B}.

To compute the radiation accurately, it is necessary to specify the momentum distribution of the particles. The conventional assumption is that the 
particles are isotropic in the jet co-moving frame, and that are injected following a certain energy distribution (typically a power-law) at \(x_0'\)\footnote{The quantities with primes refer to the jet co-moving frame.}. In the jet co-moving frame, the particle distribution function is non-stationary:
\be \label{eq:f_proper} 
f'(t',\br',\bp')=\Theta(x'-x_0'(t'))\delta(y')\delta(z')g'(p',\tau)\,, 
\ee 
where \(x_0'=x_0 /\Gamma - \beta c t'\) is the location of the acceleration site in the co-moving
frame, \(\Theta\) the Heaviside function, \(\delta\) the Dirac \(\delta\)-function, and \(\tau\) the proper age of the particle population. As non-thermal particles are accelerated at \(x'_0\),  one obtains:
\be\label{eq:tau_proper} 
\tau(t',x')=t'-\frac{x_0-\Gamma x'}{\Gamma \beta c}\,.
\ee
The evolution of the high-energy particles, which is determined by \(g'\), depends also on the parameters that define the jet orientation and \(x_0\), which for simplicity are assumed to be constant.

The internal energy of the plasma per volume unit can be obtained as:
\be\label{eq:jet_int_energy} 
e(t',\br')=
\Theta(x'-x_0'(t'))\delta(y')\delta(z') \int d^3{\mathbf p'}\, \ve' g'(p',\tau)\,,
\ee
and the energy injected\footnote{To remain consistent with the standard one-zone modeling, here one ignores the pressure of non-thermal particles; accounting for their pressure would increase the required injected luminosity by a \(\approx 30\)\%, for the same \(g\).} in non-thermal particles at \(x=x_0\) is:
\be \label{eq:normalization}
L_{\textsc{nt}} = \Gamma^2 c\beta \int d^3{\mathbf p'}\,\ve' g'(p',0)\,.
\ee

The up-scattering rate of target photons to gamma rays is determined by the differential cross-section:
\be
\frac{d\sigma}{dE_\gamma} = \frac{d\sigma}{dE_\gamma}\left({\ve,E_\gamma,\ve_{\rm ph},\theta_\textsc{sc}}\right)\,,
\ee
where \(\ve=\sqrt{m_e^2c^4+p^2c^2}\) is the electron energy, with \(m_e\) being the electron mass, \(E_\gamma\) the gamma-ray energy, \(\ve_{\rm ph}\) the target photon energy, and \(\theta_\textsc{sc}\) the scattering angle, i.e., the angle between the target photon momentum and the observer direction \citep{1981Ap&SS..79..321A}. The scattered photons move in the direction of the electron with accuracy \(\sim m_ec^2/\ve\) (\(\ll1\) in the relativistic regime), thus an observer located in the direction \({\mathbf n}_0\) should detect emission produced by particles with \(\bp=p {\mathbf n}_0\).

In the laboratory frame, the gamma-ray spectrum can be obtained from:
\be\label{eq:emission} 
\frac{dN_\gamma}{dtdE_\gamma
  d\Omega}=\int c(1-\cos\theta_\textsc{sc})\frac{d\sigma}{dE_\gamma} f(\tilde{t},{\mathbf r},p {\mathbf n}_0) p^2\, dn_{\rm ph}\, dp\, dV\,,
\ee 
where \(\theta_{\rm sc}\) in Eq.~\eqref{eq:emission} depends on \(\br\), the location of the source of target photons, and also the orbital phase, the latter playing an important role in shaping the gamma-ray lightcurve.

Photons emitted  at \({\tilde t}({\mathbf r})=T+\frac{{\mathbf r}{\mathbf n}_0}{c}\) (\(T\) is the time at which a hypothetical photon at \(\br=0\) would be produced), will be simultaneously detected by the observer.
The emission will arrive to the observer at a time \(t=T+T_\textsc{prop}\), where \(T_\textsc{prop}\) is the time required for the hypothetical photon to travel to the observer.
The linear relation between \(t\) and \(T\) implies that Eq.~\eqref{eq:emission} describes both {\it emitted} and {\it received} photons, which is not always the case \citep[see, e.g.,][]{1979rpa..book.....R}.  

The phase-space distribution function is a Lorentz invariant, \( f(t,{\mathbf r},{\mathbf p})=f'(t',{\mathbf r'},{\mathbf p'}) \), where \(t'\), \(\br'\), and \(\bp'\) are related to
\(t\), \(\br\), and \(\bp\) by Lorentz transformations \citep[e.g.,][]{1975ctf..book.....L}. For \({\tilde t}'\) and \({\tilde\tau}\) one obtains
\be\label{eq:t_proper}
  {\tilde t}'(\ti{t},x)=\Gamma T + {x\over(c\beta)}{\Db-\Gamma\over\Db\Gamma}\,,
\ee
and
\be\label{eq:tau_lab}
  {\tilde\tau}=\tau(\ti{t}',x'(\ti{t},x))={x-x_0\over\beta c \Gamma}\,,
\ee
where \(\Db=(\Gamma(1-\beta\cos\theta))^{-1}\) is the Doppler boosting factor, with \(\theta\) being  the angle between the jet velocity and the observer direction. The argument of the Heaviside function transforms simply as:
\be
\Theta\left(x'-x_0'(t')\right)=\Theta\left(\frac{x-x_0}{\Gamma}\right)=\Theta\left(x-x_0\right)\,.
\ee
Thus, one obtains that the distribution function is stationary in the laboratory frame:
\be \label{eq:f_lab}
f(t,{\mathbf r},{\mathbf p})=\Theta\left(x-x_0\right)\delta(y)\delta(z)g'\left(p',{x-x_0\over \beta c \Gamma}\right)\,,
\ee
but it is anisotropic because of the \(p_x\) dependence in \(p'=\sqrt{\Gamma(\ve/c-\beta p_x)-m_e^2c^2}\).

The integral over \(dydz\) can be computed yielding:
\be\label{eq:spectrum}
  {dN_\gamma\over dtdE_\gamma d\Omega}=\int\limits_{x_0}^{\infty}  c(1-\cos\theta_\textsc{sc}){d\sigma\over dE_\gamma}{dN_e\over d\ve dx}\, d\ve\, dn_{\rm ph}\,dx\,,
\ee
where 
\be \label{eq:effective_density}
{dN_e(\ve,x)\over d\ve dx}=g'\left({p\over\Db},{x-x_0\over \beta c \Gamma}\right) {p^2\over c}
\ee
is the effective energy distribution density\footnote{We distinguish between energy distribution, i.e., \(dN = n_1 d\ve\), and  energy distribution density, i.e., \(dN = n_2 d\ve dV\) or \(dN=n_3d\ve dx\).} of electrons emitting towards the observer. 

For the derivation of Eqs.~(\ref{eq:spectrum}) and (\ref{eq:effective_density}) one considered that the injected particle distribution is relativistic, \(\ve\simeq cp\gg \Db m_ec^2\).

Since the target density and scattering angle change along the jet, to obtain the IC emission it is necessary to know the distribution of the particles along the jet. 

\section{Evolution of non-thermal particles}
\subsection{General case}
Since in the jet co-moving frame particles are isotropic, it is convenient to use the energy distribution density:
\be \label{eq:energy_distribution}
dN=n'(\ve')d\ve' dV'=4\pi p'^2 f' dp' dV'\,.
\ee
Using the energy-momentum relation, \(\ve'^2=c^2p'^2+m_e^2c^4\), one can substitute energy as \(d\ve'=c^2p'dp'/\ve'\). Thus, one obtains \(n'=(4\pi/c^2) \ve' p' f'\). For \(x'>x'_0\),  the energy dependence of the \(g\) function can be obtained as
\be\label{eq:g_n_relation_0}
g'(p',\tau)=\int\limits_{\rm jet}^{}dy'dz'f'\propto \frac{c^2n'(\ve')}{4\pi\ve'p'}\,.
\ee
For the sake of simplicity, we normalize the particle energy density with Eq.~\eqref{eq:normalization}, i.e., with a condition that directly determines 
\(g'\). 

Under the continuous-loss approximation, particle evolution is described as:
\be \label{eq:cl_approximation}
n'(\ve')d\ve' dV'=n'(\ve_0')d\ve_0'dV'_0\,.
\ee
The energy \(\ve'_0\) in Eq.~(\ref{eq:cl_approximation}) evolves accordingly to the particle cooling equation:
\be \label{eq:cooling}
{d \ve'_0\over d\tau}=\dot{\ve}(\ve'_0,\tau)\,,
\ee
with the initial condition \(\ve'_0({\tilde\tau})=\ve'\). The non-thermal energy loss term typically accounts for synchrotron, IC, and adiabatic losses of electrons:
\be \label{eq:losses}
\dot{\ve}(\ve'_0,\tau)=\dot{\ve}_{\textsc{ic}}(\ve'_0,\tau)+\dot{\ve}_{\textsc{syn}}(\ve'_0,\tau)+\dot{\ve}_{\textsc{ad}}(\ve'_0,\tau)\,
\ee
in the jet co-moving frame.  {If the particle distribution function is defined accordingly to Eq.~(\ref{eq:f_proper}), the volume element occupied by the non-thermal particles can be taken as constant \(dV'=dV'_0\) for a constant-velocity jet (this remains correct independently of the jet geometry -see the discussion after Eq.~\ref{eq:losses_ad}-)}. Thus, one obtains:
\be\label{eq:g_n_relation}
g\propto \frac{c^2n'_0(\ve'_0)}{4\pi\ve'p'}\frac{d\ve'_0}{d\ve'}\,.
\ee

If the injection spectrum is a power-law in energy with index \(\alpha_{\textsc{inj}}\), then in the ultrarelativistic limit one obtains:
\be \label{eq:g_fun_gen}
g'(p',\tau) ={A' c^2 \ve_0'^{-\alpha_\textsc{inj}}\over 4\pi p'\ve'}  {d \ve_0' \over d\ve'}\,, 
\ee
where \(\ve_0'\) corresponds to \(\ve_0'=\ve'_0(0)\). 
The normalization constant \(A'\) is determined by Eq.~\eqref{eq:normalization}.
Then the energy distribution density, Eq.~\eqref{eq:effective_density}, can be represented as
\be \label{eq:effective_density_gen}
{dN_e(\ve,x)\over d\ve dx}={A'\Db^2\ve_0'^{-\alpha_\textsc{inj}}\over 4\pi}  {d \ve_0' \over d\ve'}
\ee
for \(\ve'=\ve/\Db\) assuming the relativistic regime, \(\ve'\approx p'c\).

\subsection{Compact emitter}
When particles are advected along the jet, the intensity of the magnetic and the photon field, and the rate of adiabatic losses, can change, meaning that the above equation does not have an analytic solution in the general case. If advection is slow compared to radiative cooling, Eq.~(\ref{eq:losses}) becomes simpler since the loss rates can be considered steady. In this case, the formal solution of Eq.~(\ref{eq:cooling}) is 
\be
\int\limits_{\ve'_0}^{\ve'}{d\hat \ve\over \dot{\ve}(\hat\ve)}=\tilde\tau\,,
\ee
which should be considered an algebraic equation that determines \(\ve'_0\) as a function of \(\ve'\) and \(\tilde\tau\). Once the original electron energy is obtained, one can express the ratio of the infinitesimal energy intervals as
\be
{d \ve_0' \over d\ve'} = {\dot\ve(\ve_0') \over \dot\ve(\ve')}\,.
\ee

\subsection{Synchrotron-Thomson losses; extended emitter}

The cooling rate for electrons interacting with background photons in the Klein-Nishina regime has a rather complicated dependence on electron energy, which makes finding an analytic solution for Eq.~(\ref{eq:cooling}) difficult.  In contrast, if the IC cooling proceeds in the Thomson regime, the energy loss term has a simple dependence on energy, which is identical to that of synchrotron cooling.

One can find an analytic solution if the dominant radiative cooling process is IC in the Thomson regime, or synchrotron emission. In this case, radiation losses have the following dependence on energy:
\be \label{eq:thomson_cooling}
\dot{\ve}_\textsc{ic}+\dot{\ve}_\textsc{syn}=-a\ve'^2\,, 
\ee
where \(a\)  does not depend on electron energy. The coefficient $a$ accounts for the synchrotron and IC losses and can vary within the jet, i.e., can be a function of \(\tau\):
\be \label{eq:cooling_coef}
\begin{split}
  a=&\,a_\textsc{ic}+a_\textsc{syn}\,,\\
  a_\textsc{ic}=&\,\frac43{\sigma_\textsc{t}c\over (m_ec^2)^2}w'_{\rm ph}\,,\\
  a_\textsc{syn}=&\,\frac43{\sigma_\textsc{t}c\over (m_ec^2)^2}w'_{\textsc{b}}\,,
\end{split}
\ee
where \(\sigma_\textsc{t}\) is the Thomson cross-section;
\(w'_{\rm ph}\) and \(w'_{\textsc{b}}\) are the energy densities of the target photons and the magnetic field in the jet frame, respectively. 

If the dominant photon field is provided by the companion star, then:
\be\label{eq:external_compton}
w'_{\rm ph}= \Db_*^{-2} {L_*\over 4\pi R^2 c}\,,
\ee
where \(L_*\) and \(R\) are the luminosity of the star and the distance to it, respectively, and the factor
\be \label{eq:star_df}
\Db_*={1\over \Gamma(1-\beta \cos\chi)}
\ee
accounts for the transformation of the photon field to the jet co-moving frame \citep[see, e.g.,][]{2012MNRAS.421.2956Z,2014ApJ...783..100K}, where \(\chi\) is the angle between the jet bulk velocity and the target photon momentum in the laboratory frame (see Fig.~\ref{fig:geom}, where the model geometry is illustrated).

The energy density of the magnetic field is:
\be
w'_{\textsc{b}}={B'^2\over 8\pi}\,,
\ee
where \(B'\) is the strength of the magnetic field in the jet co-moving frame in G.

The rate of adiabatic losses is:
\be\label{eq:losses_ad}
\dot{\ve}_{\textsc{ad}} (\ve_0')= \frac13{d\ln \rho \over d\tau}\ve_0'\,,
\ee
where \(\rho\) is the plasma density in the flow co-moving frame. We note that \(\rho\) corresponds to the jet
material. The contribution of the non-thermal particles described by Eq.~(\ref{eq:f_proper}) to this density might be
very small. The shape of the jet (e.g., cylindrical or conical) determines the rate of adiabatic losses.  The
factor \(\delta(y)\delta(z)\) in Eq.~(\ref{eq:f_lab}) does not imply any limitation on the jet shape. The meaning of
this factor is that the conditions for the non-thermal particles do not change considerably across the jet, and
Eqs.~(\ref{eq:f_proper},~\ref{eq:f_lab}) describe the properties of the particle distribution integrated over the
jet cross-section.

Equation~(\ref{eq:losses}) can be written as: 
\be
\frac d{d\tau}\left({\rho^{1/3}\over \ve'_0}\right)=\rho^{1/3}(\tau)a(\tau)\,,
\ee
which provides a relation between particle energy at the injection and the emission points as
\be\label{eq:energy_cooling_general}
\begin{split}
\left({\rho^{1/3}(\ti{\tau})\over \ve'}\right)-\left({\rho_0^{1/3}\over \ve'_0}\right)&=\int\limits_0^{\ti{\tau}} \rho^{1/3}(t')a(t')dt'\\
&=\int\limits_{x_0}^x \rho^{1/3}(\hat{x})a(\hat{x}){d\hat{x}\over \Gamma\beta c}\,,
\end{split}
\ee
where the relation between \(\ti{\tau}\) and \(x\) comes from Eq.~\eqref{eq:tau_lab}. Since the rhs of Eq.~\eqref{eq:energy_cooling_general} does not depend on energy, the ratio of the infinitesimal energy intervals is
\be\label{eq:energy_intervals_syn_thomson}
\frac{d\ve_0'}{d\ve'} = \left(\frac{\ve_0'}{\ve'}\right)^2\left(\frac{\rho(\tilde\tau)}{\rho_0}\right)^{1/3}\,,
\ee
where \(\rho_0\) and \(\ve_0'\) are the initial plasma density and particle energy, respectively. The initial energy is
\be
\ve_0'={\ve'\left({\rho_0\over\rho(\ti{\tau})}\right)^{1/3}\over1-\ve'\int\limits_0^{\ti{\tau}} a(t') \left({\rho(t')\over\rho(\ti{\tau})}\right)^{1/3}dt'}\,.
\ee

For a powerlaw injection spectrum, equation~\eqref{eq:energy_intervals_syn_thomson} allows us to obtain the electron energy distribution density with Eq.~(\ref{eq:effective_density_gen}). For a non-powerlaw injection, one should use Eq.~(\ref{eq:g_n_relation}), or equivalently Eq.~(\ref{eq:g_n_relation_0}) with the following particle energy distribution:
\be\label{eq:density_ad}
n'(\ve')={n_0'(\ve_0')\left({\rho_0\over\rho(\ti{\tau})}\right)^{1/3}\over\left(1-\ve'\int\limits_0^{\ti{\tau}} a(t') \left({\rho(t')\over\rho(\ti{\tau})}\right)^{1/3}dt'\right)^2}\,,
\ee
where \(n_0'\) is proportional to the injection spectrum.

\subsection{Synchrotron-Thomson losses; compact emitter}

If the rate of radiative losses remains constant over the cooling
distance, one can take the parameter \(a\) as a constant. However, the plasma density
dependence should be preserved in this equation, as otherwise a constant density
would imply to ignore adiabatic losses completely. The density evolution is determined by the structure of the jet. For example, for a steady conical jet the mass conservation yields:
\be \label{eq:mass_conservation}
x^2\rho \Gamma \beta = \rm const\,.
\ee

{The evolution of the macroscopic quantities in jets is a subject for dedicated (magneto)hydrodynamic simulations \citep[see, e.g.,][]{2010A&A...512L...4P}, and are beyond the scope of this paper. Equation~\eqref{eq:mass_conservation} does not allow determining the density of the jet material, since the jet may undergo bulk acceleration. For the sake of simplicity, we assume that the jet velocity remains constant in the region relevant for gamma-ray production.  In this case,}  \(\rho\) decreases as \(\propto x^{-2}\) and the integral term in Eq.~\eqref{eq:density_ad} is:
\be
\ve'a\int\limits_{x_0}^x \left({x'\over x}\right)^{-2/3}{dx'\over \Gamma\beta c} ={3a\ve'\over \Gamma\beta c}\left({x -{x_0} \left(\frac x{x_0}\right)^\frac23}\right)= 
\ee
\[
3a\ve'\left(\ti{\tau} + {x_0\over \Gamma\beta c}\left(1-\left(1+\frac {\ti{\tau}\Gamma\beta c}{x_0}\right)^\frac23\right)\right) \,.
\]
For a short cooling distance as compared to \(x_0\), this reduces to:
\be
\ve'a\int\limits_{x_0}^x \left({x'\over x}\right)^{-2/3}{dx'\over \Gamma\beta c}\approx {a\ve'\ti{\tau}}\,, 
\ee
which coincides with a solution without adiabatic losses. This calculation illustrates, to some extent, a trivial physical fact: in steady jets the adiabatic losses might be important only in extended emitters. 

As shown above, assuming a compact emitter (CE) in a steady jet implies a small impact of adiabatic losses. Adopting a power-law in energy with index \(\alpha_\textsc{inj}\) for the injected particles, one obtains for a compact emitter:
\be\label{eq:n_prime}
n'(\ve')\propto(1-a\ti{\tau}\ve')^{\alpha_\textsc{inj}-2}\ve'{}^{-\alpha_\textsc{inj}}\,.
\ee
Accordingly with Eq.~\eqref{eq:g_fun_gen}, one obtains that
\be \label{eq:g_primed}
g'\left({p\over\Db},{x-x_0\over \beta c \Gamma}\right)=\frac{A'\Db^2c}{4\pi p^2}\left(1-a{x-x_0\over \beta c \Gamma}{pc\over\Db}\right)^{\alpha_\textsc{inj}-2}\left({cp\over\Db}\right)^{-\alpha_\textsc{inj}},
\ee
in the limit of \(\ve\simeq cp\gg \Db m_e c^2\). The maximum energy in Eq.~(\ref{eq:g_primed}), \(\ve_\textsc{max}'\), is limited by the injection process, and the following relation should be
fulfilled: \(\ve'/(1-a\ti{\tau}\ve')<\ve_\textsc{max}'\).  In Eq.~\eqref{eq:spectrum}, the limit imposed by the maximum energy translates into the integral upper limit:
\be\label{eq:x_max}
x_\textsc{max}=x_0+{c\beta\Gamma\over a}{\ve_\textsc{max}'-\ve'\over \ve_\textsc{max}'\ve'}\,.
\ee

The dominance of radiation cooling over advection implies that the cooling length, \((x_\textsc{max}-x_0)\), is small as compared to the characteristic distance over which the loss rate can change significantly. Under these conditions, as noted, one can compute the integral over \(x\) analytically:
\[
\int\limits_{x_0}^{x_\textsc{max}}dx\left(1-a{x-x_0\over \beta \Gamma}{p\over\Db}\right)^{\alpha_\textsc{inj}-2}=
\]
\be \label{eq:x_integral}
\quad\quad\frac{\beta \Db\Gamma}{ap(\alpha_\textsc{inj}-1)}\left[1-\left(1-u_\textsc{max}\right)^{\alpha_\textsc{inj}-1}\right]
\simeq\frac{\beta \Db\Gamma}{ap(\alpha_\textsc{inj}-1)}\,,
\ee
where \(u_\textsc{max}=(\ve'_\textsc{max}-cp/\Db)/\ve'_\textsc{max}\simeq1\), for a large maximum injection energy.

\section{Inverse Compton Emission from a Compact relativistic Emitter}

Combining Eqs.~(\ref{eq:spectrum}), (\ref{eq:effective_density}), (\ref{eq:g_primed}), and (\ref{eq:x_integral}), one obtains the following expression for the spectrum produced in a compact gamma-ray emitter in the case of dominant Thomson or synchrotron losses:
\be
{dN_\gamma\over dtdE_\gamma d\Omega}\simeq 
\ee
\[
\quad A'\int c(1-\cos\theta_\textsc{sc}){d\sigma\over dE_\gamma} \frac{\beta \Db\Gamma}{ap(\alpha_\textsc{inj}-1)}\frac{\Db^2c}{4\pi p^2}\left({cp\over\Db}\right)^{-\alpha_\textsc{inj}}  p^2\, dp\, dn_{\rm ph}\,.
\]
Accordingly to Eq.~(\ref{eq:normalization}) for a power-law injection with \(\alpha_\textsc{inj}>2\), the normalization coefficient \(A'\) is approximately
\be
\label{eq:norm}
A'\approx {L_\textsc{nt}(\alpha_\textsc{inj}-2)\over \Gamma^2\beta }\ve_\textsc{min}^{\alpha_\textsc{inj}-2}\,,
\ee
yielding: 
\[
  {dN_\gamma\over dtdE_\gamma d\Omega}=\quad{L_\textsc{nt}\ve^{\alpha_\textsc{inj}-2}_\textsc{min}c\over 4\pi a}{\alpha_\textsc{inj}-2\over\alpha_\textsc{inj}-1} {\Db^{3+\alpha_\textsc{inj}}\over \Gamma} \times
\]
\be
\int c(1-\cos\theta_\textsc{sc}){d\sigma\over dE_\gamma} \ve^{-(\alpha_\textsc{inj}+1)}\, d\ve\, dn_{\rm ph}\,.
\ee
The relativistic limit, \(\ve\approx pc\), was used above for the sake of simplicity.

{If IC scattering of stellar photons is the dominant cooling channel, \(a\approx a_\textsc{ic}\), where \(a_\textsc{ic}\) is defined by Eqs.~(\ref{eq:cooling_coef}) and (\ref{eq:external_compton}),} one obtains:
\be\label{eq:spectrum_compact}
  {dN_\gamma\over dtdE_\gamma d\Omega}=\left[{\Db^{2\alpha_\gamma+1}\Db_*^2\over \Gamma} \right]\int c(1-\cos\theta_\textsc{sc}){d\sigma\over dE_\gamma} {d\tilde{N}\over d\ve}\, d\ve\, dn_\textsc{bb}\,,
\ee
where \(n_\textsc{bb}=(2R/R_*)^2n_{\rm ph}\) is the Planck distribution, \(R_*\) the radius of the optical star, and the function \(\tilde{N}\) depends on the basic parameters characterizing the system (stellar temperature \(T_*\) and available non-thermal power), and the acceleration mechanism (accelerated particle energy dependence and minimum energy):
\be 
\label{eq:n_0}
{d\tilde{N}\over d\ve}=\frac3{64\pi} {L_\textsc{nt}\over\sigma_\textsc{b} \sigma_\textsc{t} T_*^4m_e c^2}\left({\ve_\textsc{min}\over m_ec^2}\right)^{-3}{\alpha_\textsc{inj}-2\over\alpha_\textsc{inj}-1} \left({\ve\over \ve_\textsc{min}}\right)^{-(\alpha_\textsc{inj}+1)}\,,
\ee
where \(\sigma_\textsc{b}\) is the Stefan-Boltzmann constant, and the gamma-ray photon index is related to \(\alpha_\textsc{inj}\) through
\(\alpha_\textsc{inj}=2(\alpha_\gamma-1)\). 

If the injection process remains steady over a time similar to, or longer than, the orbital period, there are two factors that affect the variability of the GeV gamma-ray emission.  The changing scattering angle and the relativistic effects can vary with the orbital phase. The relativistic effects are accounted by the term \(\left[{\Db^{2\alpha_\gamma+1}\Db_*^2\Gamma^{-1} }\right]\), which includes both Doppler boosting of the emission, and the transformation of the stellar photon field to the jet RF. 

In line with Eq.~\eqref{eq:effective_density}, one can also consider the effective energy distribution of particles in the whole jet for a compact emitter:
\be\label{eq:n_eff}
{dN_e^{(\textsc{ce})}\over d\ve}= \int\limits_{x_0}^\infty dx\,{dN_e(\ve,x)\over d\ve dx}=
\ee
\[
\left[{\Db^{2\alpha_\gamma+1}\Db_*^2\over \Gamma} \right]\frac3{4} {L_\textsc{nt} R^2\over L_* \sigma_\textsc{t}m_e c^2}\left({\ve_\textsc{min}\over m_ec^2}\right)^{-3}{\alpha_\textsc{inj}-2\over\alpha_\textsc{inj}-1} \left({\ve\over \ve_\textsc{min}}\right)^{-(\alpha_\textsc{inj}+1)}\,.
\]
The emission spectrum can be written as:
\be\label{eq:spectrum_compact_v2}
  {dN_\gamma\over dtdE_\gamma d\Omega}=\int c(1-\cos\theta_\textsc{sc}){d\sigma\over dE_\gamma} {d N_e^{(\textsc{ce})}\over d\ve}\, d\ve\, dn_{\rm ph}\,,
\ee
where the target photon density contains the dilution factor
\be\label{eq:dilution}
n_{\rm ph}=\left({R_*\over 2R}\right)^2n_\textsc{bb}\,.
\ee

To model the GeV emission from \cyg, \citet{2010MNRAS.404L..55D} introduced a factor related to relativistic effects obtained for a blob emitter. \citet{2012MNRAS.421.2956Z} argued that this enhancement factor is not applicable to jet sources, and proposed instead the factor derived by \citet{1997ApJ...484..108S} for the enhancement of the emission in blazars. The impact of stellar field relativistic boosting, \(\Db_*\), is also accounted in the model considered by \citet{2012MNRAS.421.2956Z}. In that study, this factor affects several different parameters: electron density; scattering angle; and scattering rate. However, combining their Eqs.~(15), (22), and (A9), and taking into account that \(x\) (in the notation of \citealt{2012MNRAS.421.2956Z}) is determined by the scattering angle in the jet frame (\(x=\Db\Db_*(1-\cos\theta)\), where \(\theta\) is the scattering angle in the laboratory frame), one can derive that the IC flux is indeed \(\propto\Db_*^2\), which agrees with Eqs.~(\ref{eq:n_eff}), and (\ref{eq:spectrum_compact_v2}).

For a power-law energy distribution of electrons, the gamma-ray spectrum in the Thompson regime allows a simple analytic approximation \citep{2010MNRAS.404L..55D,2012MNRAS.421.2956Z}. This allows us to obtain the dependence of the IC flux analytically
\citep[for the exact expression see][]{2012MNRAS.421.2956Z}:
\be\label{eq:spectrum_compact_v3}
  {dN_\gamma\over dtdE_\gamma d\Omega}\propto(1-\cos\theta_\textsc{sc})^{\alpha_\textsc{inj}+2\over2} E_\gamma^{-{\alpha_\textsc{inj}+2\over2}}\,.
\ee
 
\section{Evolution of non-thermal electrons in an Extended Emitter}

Equation~\eqref{eq:spectrum_compact} describes the IC emission if the cooling length is short as compared to \(x_0\) (see Eq.~\ref{eq:x_max}) and the scattering proceeds in the Thomson regime. If \(\ve\) is the energy responsible for the generation of gamma rays in the laboratory frame, then combining Eqs.~\eqref{eq:cooling_coef} and \eqref{eq:x_max} one obtains the condition for a compact production site:
\be
{c\beta\Gamma\Db\Db_*^2\over a_0 x_0\ve}\left({R\over d}\right)^2\ll1\,,
\ee
with
\be \label{eq:cooling_coef0}
a_0=\frac43{\sigma_\textsc{t}\over (m_ec^2)^2}{L_*\over 4\pi d^2 }\simeq {L_{39}\over 6d_{12}^2}\rm\,GeV^{-1}\,s^{-1}\,,
\ee
where \(d\) is the separation between the normal star and the CO, and fiducial parameter values between those of the gamma-ray emitting microquasars \cyg and \cygl were used: \(L_*=10^{39}L_{39}\,\ergs\) and \(d=10^{12}d_{12}\rm\,cm\). At typical \fer energies, say \(E_\gamma\sim200\rm\,MeV\),
the dominant contribution is produced by electron energies of a few GeV (\(\ve=\hat\ve\rm \,GeV\)). Thus, the condition for the applicability of the compact emitter approximation becomes:
\be
0.2{\beta\Gamma\Db\Db_*^2d_{12}\over \hat{\ve}L_{39}}\left({x_0\over d}\right)^{-1}\left({R\over d}\right)^2\ll1\,,
\ee
and can be violated even for a mildly relativistic jet with \(\Gamma\sim 2\). 

It is worth noting that for higher energy electrons, IC scattering proceeds in the Klein-Nishina regime. This should lead to an even larger extension of the production site, since IC losses in the Klein-Nishina regime proceed slower than in Thomson. 

Losses through IC are not necessarily the dominant cooling mechanism. As indicated above, synchrotron and adiabatic cooling can be also considered. Since adiabatic losses trace a change in the density, they are to be included when the compact emitter approximation fails.  

If synchrotron losses dominate, the condition for a compact emitter is determined by the magnetic field in the jet. The synchrotron time is
\be
t'_\textsc{syn}\simeq 4\times10^5 \left({B'\over \rm\,G}\right)^{-2}\left({\hat{\ve}\over\Db}\right)^{-1}\rm s\,.
\ee
  The cooling length, \(x_\textsc{syn}=\Gamma\beta ct'_\textsc{syn}\) remains small as compared to \(x_0\) if the magnetic field is:
\be\label{eq:b_dominant}
B'\gg 100 \sqrt{\beta \Gamma\Db \over \hat\ve x_{0,12}} \rm\,G\,.
\ee
For the sake of simplicity, let us parameterize the jet radius as a fraction of \(x_0\): \(r_{\rm j}=\varpi x_0\), where \(\varpi \ll1\). Then the Poynting energy flux in such a jet is 
\be
\begin{split}
S&=\frac {VB'^2 \Gamma^2\varpi^2x_0^2}{4\pi} \\
&\gg 2.5\times10^{37} {\varpi^2\beta^2\Db\Gamma^3\over \hat\ve} \left({x_0\over 10^{12}\rm\,cm}\right)^{3/2}\ergs\,.
\end{split}
\ee
The numerical coefficient \(2.5\times10^{37}\) corresponds to approximately \(5\%\) of the Eddington luminosity for a \(5M_\odot\) black
hole.  Although such a strong magnetic field cannot be excluded in Galactic jet sources, modeling tends to favour a weaker
magnetization of the jet \citep{2010MNRAS.404L..55D,2012MNRAS.421.2956Z,2018MNRAS.479.4399Z}. Thus, we conclude that there are no robust arguments excluding a significant extension of the gamma-ray production region in gamma-ray emitting microquasars.

\subsection{Synchrotron and adiabatic loss treatment}

The structure of Eq.~\eqref{eq:energy_cooling_general} allows us to consider IC losses and synchrotron losses independently. First we will start with synchrotron losses. The combined impact of synchrotron and adiabatic losses is determined by the following integral:
\be
\int\limits_0^{\ti{\tau}} \rho^{1/3}(t')a(t')dt'\propto\int\limits_{x_0}^{x} \rho^{1/3}(\hat{x})B'^2(\hat{x}){d\hat{x}}\,.
\ee

The density and the magnetic field depend on the structure of the jet, but most likely they can be approximated by power-law functions in a limited section of the jet. Thus, the above integral can be solved analytically yielding a description of the joint impact of synchrotron and adiabatic losses. For example, if the impact of  IC losses is small, then such a treatment of adiabatic and synchrotron losses allows us to obtain a solution for Eq.~(\ref{eq:energy_cooling_general}), and thus to obtain an analytical description of the spatial-energy distribution of electrons in the jet.

\subsection{Thomson and adiabatic loss treatment}

For the case of dominant IC losses, the function \(a\) has a more complicated structure since it depends not only on the distance from the CO, but also on the angle between the jet velocity and target photon momenta. 
Modeling by \citet{2010MNRAS.404L..55D} suggests that the jet in \cyg is not perpendicular to the orbital plane. Thus, we consider here also a case when the jet is inclined by a fixed angle \(\alpha\). In this case, the cooling rate depends
on the parameter \(\zeta=(x/d-\cos\alpha)\), where
\(\alpha\) is the angle between the jet velocity and the direction
from the CO to the companion star  (see Fig.~\ref{fig:geom}, where the model geometry is illustrated). Including \(\zeta\),
Eq.~\eqref{eq:cooling} describing particle cooling becomes:
\be\label{eq:cooling_extended}
\frac d{d\zeta}{\rho^{1/3}\over \ti{\ve}}=\rho^{1/3}(x(\zeta)){\zeta^2\left(1+\beta^2\right)+\sin^2\alpha -2\beta\zeta\sqrt{\zeta^2+\sin^2\alpha}\over\left(\zeta^2+\sin^2\alpha\right)^2}\,,
\ee
where the dimentionless energy, \(\ti{\ve}\), is defined as 
\be
\ti{\ve}=\ve'_0{L_*\Gamma\sigma_\textsc{t}\over 3\pi c\beta d (m_ec^2)^2}\,.
\ee
In general, Eq.~\eqref{eq:cooling_extended} does not allow an analytic solution. We can still treat it numerically, for which we will adopt a density profile of \(\rho= \rho_0(x_0/x)^{2}\). 

\subsection{Thomson loss treatment}

In case adiabatic losses are weak, i.e., \(\rho\) is roughly constant, Eq.~\eqref{eq:cooling_extended} determines \(\ti{\ve}(x)\):
\be\label{eq:solution_extended_0}
{1\over \ti{\ve}}={\pi/2 - \chi\over \sin\alpha}\left(1+\frac{\beta^2}2\right)+{2\beta d\over R}-{\beta^2 d\sqrt{R^2 - d^2\sin^2\alpha} \over 2R^2} + {\rm C}\,,
\ee
where \(\chi=\chi({x})\) and \(R=R({x})\) are functions of \({x}\), and
\(\rm C\) is an integration constant. This relation allows one to link the particle energy
at the injection point \(x_0\) to that at the emission point \(x\):
\be\label{eq:solution_extended}
{1\over \ti{\ve}_0}+{E(x_0)} = {1\over \ti{\ve}}+{E(x)}\,,
\ee
where
\be
{E(x)}=-{\frac\pi2-\chi(x)\over \sin\alpha}\left(1+\frac{\beta^2}2\right)-{2\beta d\over R(x)}+{\beta^2 d\sqrt{R(x)^2 - d^2\sin^2\alpha} \over 2R(x)^2}\,.
\ee

Equation~\eqref{eq:solution_extended} together with Eq.~\eqref{eq:cl_approximation} allow us to obtain an analytic representation of the energy distribution of the particles in the jet:
\be \label{eq:extended_solution}
n'(\ve')={\ve_0'^2\over\ve'^2}n'_0(\ve_0')\,,
\ee
where the initial particle energy,
\be
\ve_0'={\ve'\over 1 - {\ve'}{L_*\Gamma\sigma_\textsc{t}(E(x_0)-E(x))\over 3\pi c\beta d (m_ec^2)^2}}\,,
\ee
should remain smaller than the maximum energy, \(\ve'_\textsc{max}\), in the injection spectrum \(n'_0\).

For a power-law injection spectrum, one obtains that the distribution function is
\[
g'\left({p\over\Db},x\right)=
\]
\be\label{eq:ext_distr}
\frac{A'\Db^2c}{4\pi p^2}\left(1-{L_*\Gamma\sigma_\textsc{t}(E(x_0)-E(x))\over 3\pi \beta d (m_ec^2)^2}{p\over\Db}\right)^{\alpha_\textsc{inj}-2}\left({cp\over\Db}\right)^{-\alpha_\textsc{inj}}\,.
\ee

Equation~\eqref{eq:ext_distr} describes the Thomson cooling of high energy
electrons in the jet also in the case when electrons may travel
distances comparable to the orbital separation. This analytic solution does not account for the impact of adiabatic losses, as it is the case for instance in a cylindrical jet. Equation~\eqref{eq:ext_distr} can be easily generalized to account for synchrotron losses, for example, if the magnetic field strength has a power-law dependence. 

For the computation of the IC emission, one should note that in Eq.~\eqref{eq:spectrum} the
photon density \(n_{\rm ph}\) and the IC scattering angle depend on \(x\). In the next section we apply a
numerical approach to compute the radiation from such an extended region.

\section{Radiation from an extended emitter}\label{sec:radiation}

To study the impact of advection in the jet we adopt parameters similar to those of \cyg, which is a highly compact system with a very bright star, and also one of the most powerful GeV sources in the Galaxy.
The temperature and luminosity of the donor star were adopted to be
\(T_*=10^5\rm\,K\) and \(L_*=1.8\times10^{39}\,\ergs\), respectively, and the CO was assumed to be located at a distance of \(d=2.7\times10^{11}\rm \,cm\) from the companion star. This separation distance is the largest allowed by \citet{2017MNRAS.472.2181K}, and corresponds to a CO with \(M_\textsc{co}\simeq5M_\odot\) and a WR star with \(M_\textsc{wr}\simeq15M_\odot\). 
\subsection{Emitter size}
First, we study the size of the gamma-ray emitting region for different locations of the
acceleration site, and for different orientations of the jet. We adopt the maximum energy in the injection spectrum to be \(\ve_\textsc{max}'=10\rm\,GeV\), and compute the maximum distance that electrons with energy \(\ve'\) can reach. For electrons with larger energy, \(\geq10\rm\,GeV\), and the stellar temperature of \cyg, the Klein-Nishina effect should weaken the IC losses as compared to the Thomson case. This should lead to an increase of the advection distance, although the impact on the GeV lightcurve should be small given the steepness of the measured spectrum. Efficient advection of high-energy electrons may have an important influence on the multi-GeV gamma-ray flux, since in this energy band the gamma-gamma attenuation is strong, and more efficient advection of the emitting electrons can significantly reduce the attenuation factor.

The advection distance, $x_\textsc{max}$, is determined through Eq.~\eqref{eq:cooling_extended} as 
\be \label{eq:x_max_ad}
{\rho^{1/3}_0\over \ti{\ve}_\textsc{max}}-{\rho^{1/3}(x_\textsc{max})\over \ti{\ve}}=
\ee
\[
\int\limits_{\zeta({x}_0)}^{\zeta(x_{\textsc{max}})} d\zeta\rho^{1/3}(x(\zeta)){\zeta^2\left(1+\beta^2\right)+\sin^2\alpha -2\beta\zeta\sqrt{\zeta^2+\sin^2\alpha}\over\left(\zeta^2+\sin^2\alpha\right)^2}\,,
\]
which in the case of weak adiabatic losses reduces to
\be\label{eq:x_max_ic}
{1\over \ti{\ve}_\textsc{max}}+{E(x_0)} = {1\over \ti{\ve}}+{E(x_\textsc{max})}\,.
\ee

\begin{figure}
  \includegraphics[width=\columnwidth]{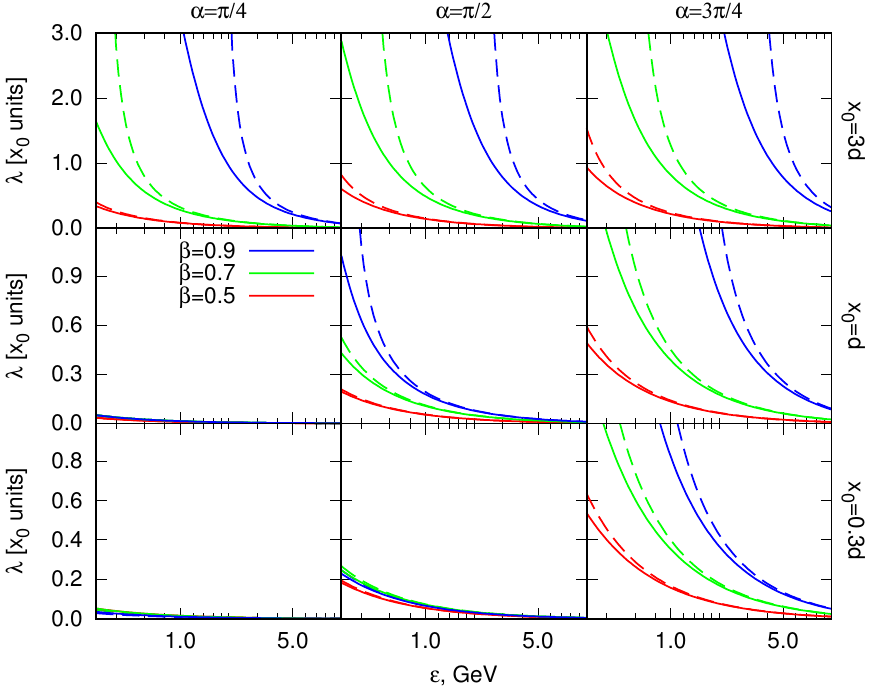}
  \caption{Relative size of the jet filled with electrons of different energies. Calculations are performed for conditions similar to \cyg, three different jet orientations (columns for \(\alpha=\pi/4\), \(\pi/2\), and \(3\pi/4\)), injection locations (raws: \(x_0=0.3d\), \(d\), and \(3d\)), and jet velocities (colors: \(\beta=0.5\) -red-, \(\beta=0.7\) -green-, and \(\beta=0.9\) -blue-). The case with negligible adiabatic losses is shown with dashed lines, and for a conical jet with solid lines.}
  \label{fig:size}
\end{figure}

We solve numerically Eq.~\eqref{eq:x_max_ad} for a conical jet and several jet velocities and orientations. However, if the jet expands slower than a conical one, the rate of adiabatic losses will be smaller. Thus, we also compute the case for dominant IC losses (Eq.~\ref{eq:x_max_ic}). Figure~\ref{fig:size} shows the relative size of the emitting region,
\be
\lambda = {x_\textsc{max}(\ve)-x_0 \over x_0}\,,
\ee
for different injection points: \(x_0=0.3 d\), \(d\), and \(3d\) (from bottom to top); jet inclinations: \(\alpha=\pi/4\), \(\pi/2\), and \(3\pi/4\) (from left to right); and jet velocities: \(\beta=0.5\), \(0.7\), and \(0.9\) (line color). It is seen that even in such a compact system as \cyg, the advection of GeV electrons might be significant if the acceleration site is located at \(d\sim10^{12}\rm cm\) from the CO, in a jet that moves with a bulk Lorentz factor \(\Gamma\geq2\) ($\beta\geq 0.87$). 

\begin{figure}
  \includegraphics[width=\columnwidth]{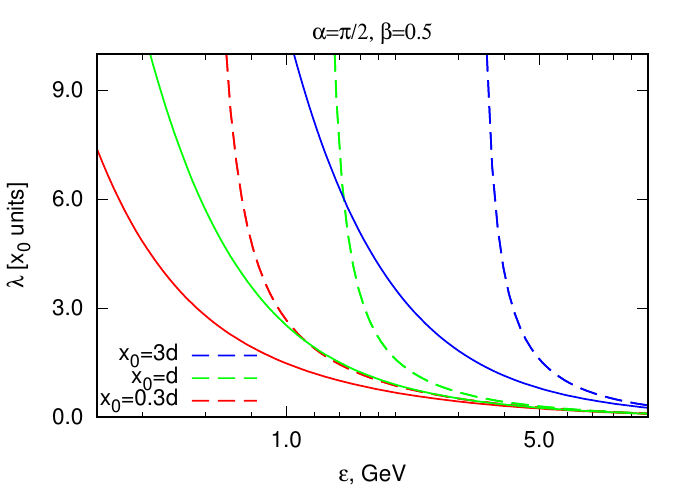}
  \caption{Relative size of the jet filled with electrons of different energies. Calculations are performed for conditions similar to those in \cygl, assuming a jet perpendicular to the orbital plane and a jet velocity of \(\beta=0.5\). Three different injection points are shown: \(x_0=0.3d\) (red), \(x_0=d\) (green), and \(x_0=3d\) (blue). The case with negligible adiabatic losses is shown with dashed lines, and adiabatic losses for a conical jet with solid lines.}
  \label{fig:size_CygX1}
\end{figure}

For a system with a larger star separation than \cyg, the impact of advection should be stronger. For parameters similar to \cygl \citep[\(T_*=3\times 10^4\rm\,K\), \(L_*=8\times10^{38}\,\ergs\), and \(d=3\times10^{12}\rm \,cm\); see, e.g.,][and reference therein]{2009ApJ...701.1895C}, we show in Fig.~\ref{fig:size_CygX1} the extension of the GeV emitter for three different injection points: \(x_0=0.3 d\), \(d\), and \(3d\). As seen from the figure, under dominant IC losses the compact emitter approximation for \cygl (or similar systems) is only justified if the injection occurs very close to the CO, or if the jet is not relativistic.

\subsection{Particle distribution}
\begin{figure}
  \includegraphics[width=\columnwidth]{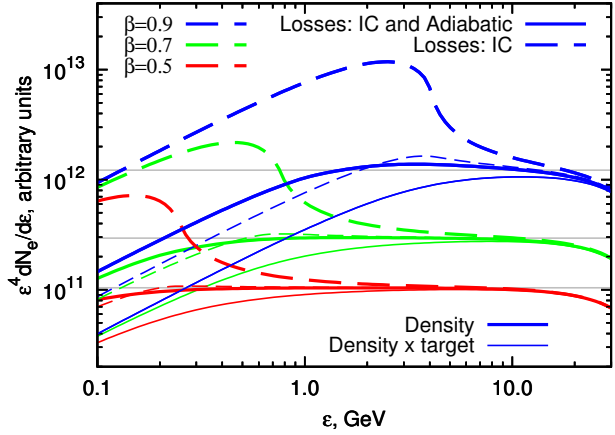}
  \caption{Energy distribution of electrons in the jet calculated for conditions similar to those in \cyg. Three different jet velocities are shown: \(\beta=0.5\) (red), \(\beta=0.7\) (green), and \(\beta=0.9\) (blue). The case with negligible adiabatic losses is shown with dashed lines, and for a conical jet with solid lines. Gray lines show the energy distribution obtained for a compact emitter, Eq.~\eqref{eq:n_eff}. Thin lines show the electron distribution weighted with the target photon density (see Eq.~\eqref{eq:density_weighted}). Other model parameters were set as \(B=0\), \(x_0=3d\), and \(\alpha=\pi/2\). }
  \label{fig:density_CygX3}
\end{figure}

 Advection affects the particle energy distribution in the jet. As the baseline case we adopt the results obtained for a compact emitter, Eq.~\eqref{eq:n_eff}. We assume that the nonthermal injection occurs at \(x=3d\) and the acceleration spectrum is steep: \(\alpha_\textsc{inj}=3\). As follows from Eq.~(\ref{eq:spectrum_compact_v3}), for this electron injection spectrum the gamma-ray photon index should be \(\sim2.5\), which is roughly consistent with the spectrum detected from \cyg with \fer \citep[see, e.g.,][]{2018MNRAS.479.4399Z}. Thomson cooling should render an energy distribution \(\propto\ve^{-4}\). Thus, in Fig.~\ref{fig:density_CygX3} we plot the particle distributions multiplied by \(\ve^4\). For the case of an extended emitter (EE), we define the effective energy distribution in accordance with Eq.~\eqref{eq:effective_density} and similarly to Eq.~\eqref{eq:n_eff} as:
\be
{dN_e^{(\textsc{ee})}(\ve)\over d\ve}= \int\limits_{x_0}^\infty dx\,{dN_e(\ve,x)\over d\ve dx}\,.
\ee

In Fig.~\ref{fig:density_CygX3}, the results of the calculations of the electron energy distribution 
are shown for \(\beta=0.5\), \(0.7\), and \(0.9\), and \(x_0=3d\). The maximum energy in the injected spectrum was assumed to be \(\ve'_\textsc{max}=30\rm\,GeV\), the Doppler boosting factor  \(\Db=1.7\) (which implies different viewing angles for different jet velocities), and the magnetic field was assumed to be weak (formally set to \(B=0\)). The energy distribution of a compact emitter, Eq.~(\ref{eq:n_eff}), for the same jet parameters, is shown with gray lines in Fig.~\ref{fig:density_CygX3}. 

As shown in Fig.~\ref{fig:density_CygX3}, the energy distributions that account for advection have three characteristic features as compared to Eq.~\eqref{eq:n_eff}.  At the highest energies, there are less particles than in the compact emitter approximation, which is caused by ignoring the injection maximum energy in  Eq.~\eqref{eq:n_eff} (the compact emitter case),  and thus it should not be associated with advection. At lower energies, advection leads to particle accumulation in regions with slower energy losses. Thus, for the same injection, the amount of particles for a steady jet is higher.  Obviously, these features are the mostly pronounced in electron distributions computed for weak adiabatic losses. Adiabatic losses expected in conical jets appear to be significant enough to determine the shape of the electron distribution at lower energies as shown in Fig.~\ref{fig:density_CygX3} with thin lines. 

To illustrate the effect of photon target dilution, we also compute the amount of emitting electrons weighted by the target photon density (thin lines in Fig.~\ref{fig:density_CygX3}):
\be \label{eq:density_weighted}
{d \bar{N_e}^{(\textsc{ee})}\over d\ve}=\int\limits_{x_0}^\infty dx\, {d N_e\over d\ve dx} \left({R_0\over R(x)}\right)^2\,.
\ee

As seen in the figure, the weighted energy distribution is suppressed with respect to the compact emitter approximation at low energies because of particle escape. In the calculations, we adopted a maximum jet length of \(35d\), as particles reaching that far from the CO are already strongly cooled down due to adiabatic losses. We also note that particles that reached that distance from the CO should not produce variable gamma-ray emission, as $\theta_{\rm sc}$ does not significantly change along the orbit. Figure~\ref{fig:density_CygX3} shows that for  \(\beta\geq0.7\), the particle energy distribution features a break or steepening at energies around \(\ve\simeq1\rm\,GeV\), which might be testable with \fer in \cyg. 

\subsection{Spectral energy distribution and gamma-ray lightcurve}

\begin{figure}
  \includegraphics[width=\columnwidth]{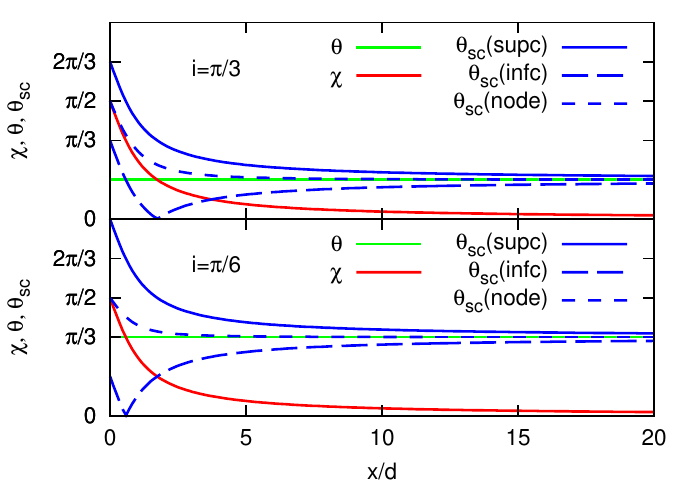}
  \caption{Characteristic angles that determine electron cooling and emission in a relativistic jet. Figure shows \(\chi\) (target-photon momentum -- bulk velocity), \(\theta\) (observer direction -- bulk velocity), and \(\theta_\textsc{sc}\) (observer direction -- target-photon momentum) angles. Calculations are performed for two orbital inclinations: \(i_\textsc{orb}=60^\circ\) (top panel) and \(i_\textsc{orb}=30^\circ\) (bottom panel). The jet was assumed to be perpendicular to the orbital plane, $\alpha=\pi/2$, and thus only \(\theta_\textsc{sc}\) depends on the orbital phase; three different orbital phases are shown: SUPC (solid lines), INFC (long dashed lines), and NODE (dashed lines).}
  \label{fig:angles}
\end{figure}

To compute the radiative output from a jet in a binary system, it is necessary to define the inclination of the orbit and the jet orientation. The jet orientation should also include the assumption if one considers jet or counter-jet; the former is located on the same side as the observer, and the latter on the opposite side, with respect to the orbital plane.  We consider two cases: \(i_\textsc{orb}=30^\circ\) and \(i_\textsc{orb}=60^\circ\), and the emission produced only by the jet since for relativistic bulk velocities the emission from the counter-jet is strongly suppressed. To illustrate the impact of advection we consider the simplest case of a circular orbit, which is reasonable for \cyg. For simplicity, the jet is assumed to be perpendicular to the orbital plane. In this case, there are three relevant orbital phases: superior/inferior conjunction of the CO (SUPC/INFC); and when the CO crosses the plane of the sky (NODE). Inverse Compton losses and radiation are affected by three angles: \(\chi\) (jet velocity-photon momentum angle); \(\theta\) (jet velocity-line of sight angle); and \(\theta_\textsc{sc}\) (IC scattering angle), which are shown in Fig.~\ref{fig:geom}. These angles depend on the inclination of the orbit, the jet orientation, and the emitter location in the jet. The dependence of these angles on the distance from the CO is shown in Fig.~\ref{fig:angles} for the selected inclinations.

\begin{figure}
  \includegraphics[width=\columnwidth]{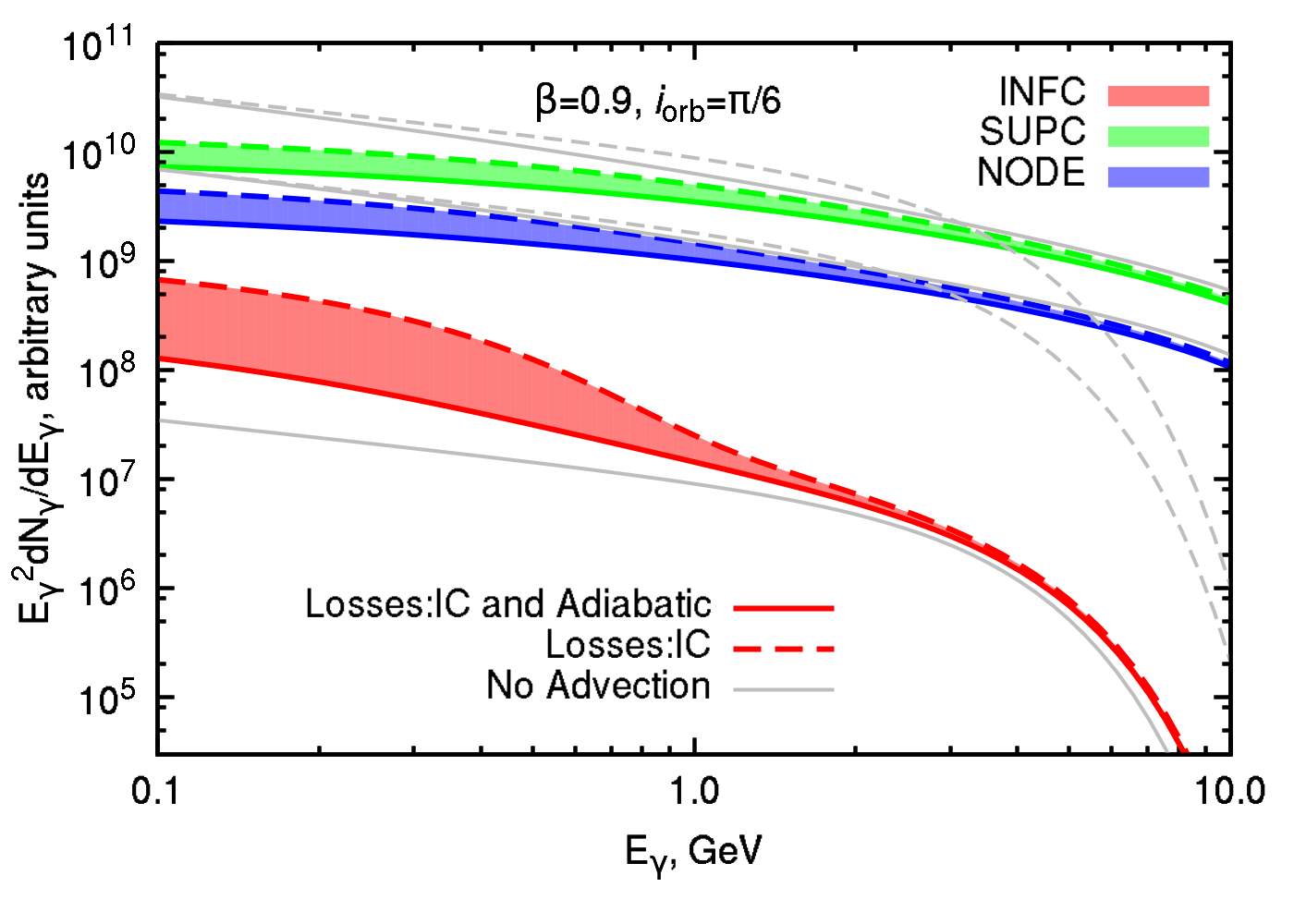}
  \includegraphics[width=\columnwidth]{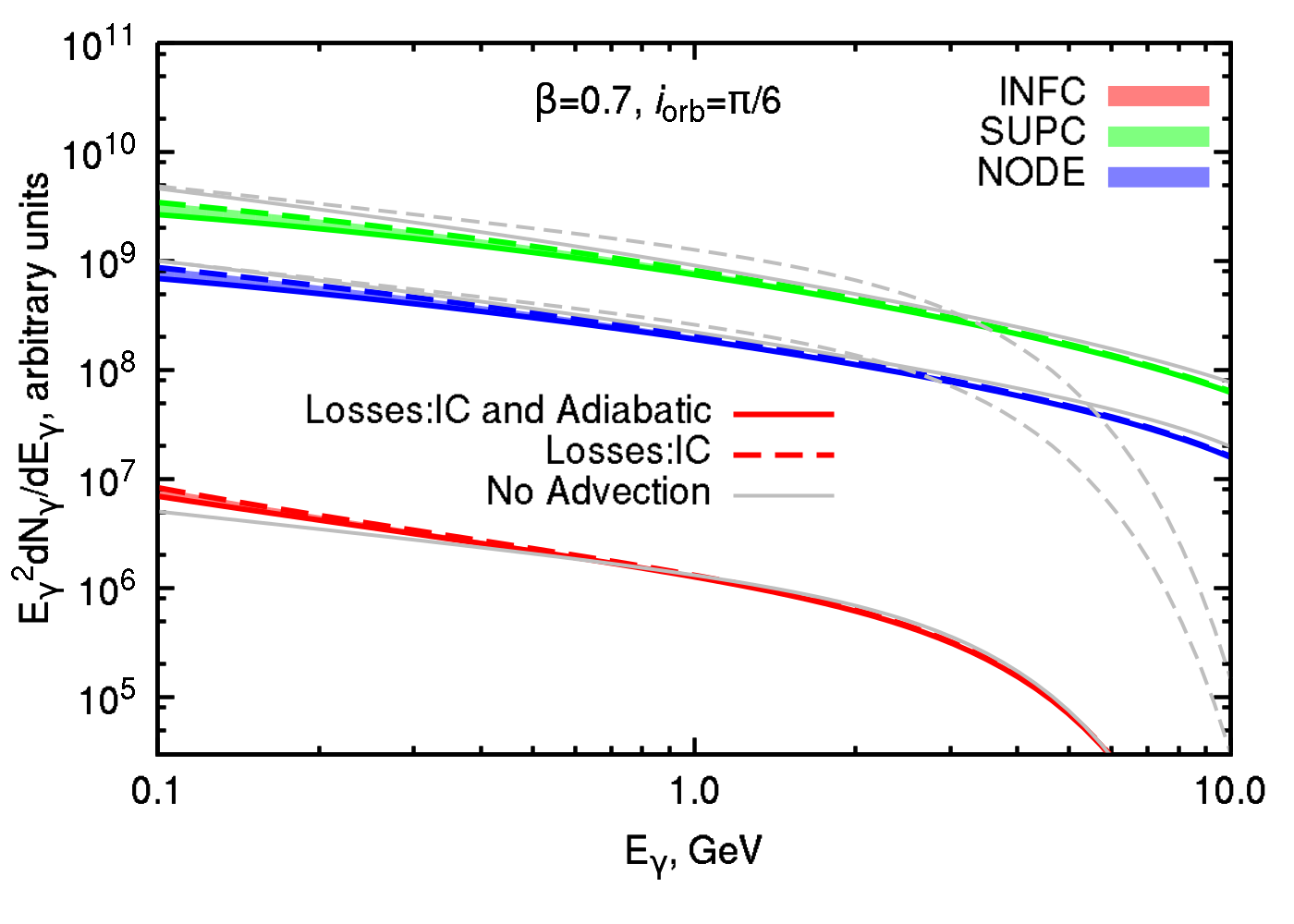}
  \includegraphics[width=\columnwidth]{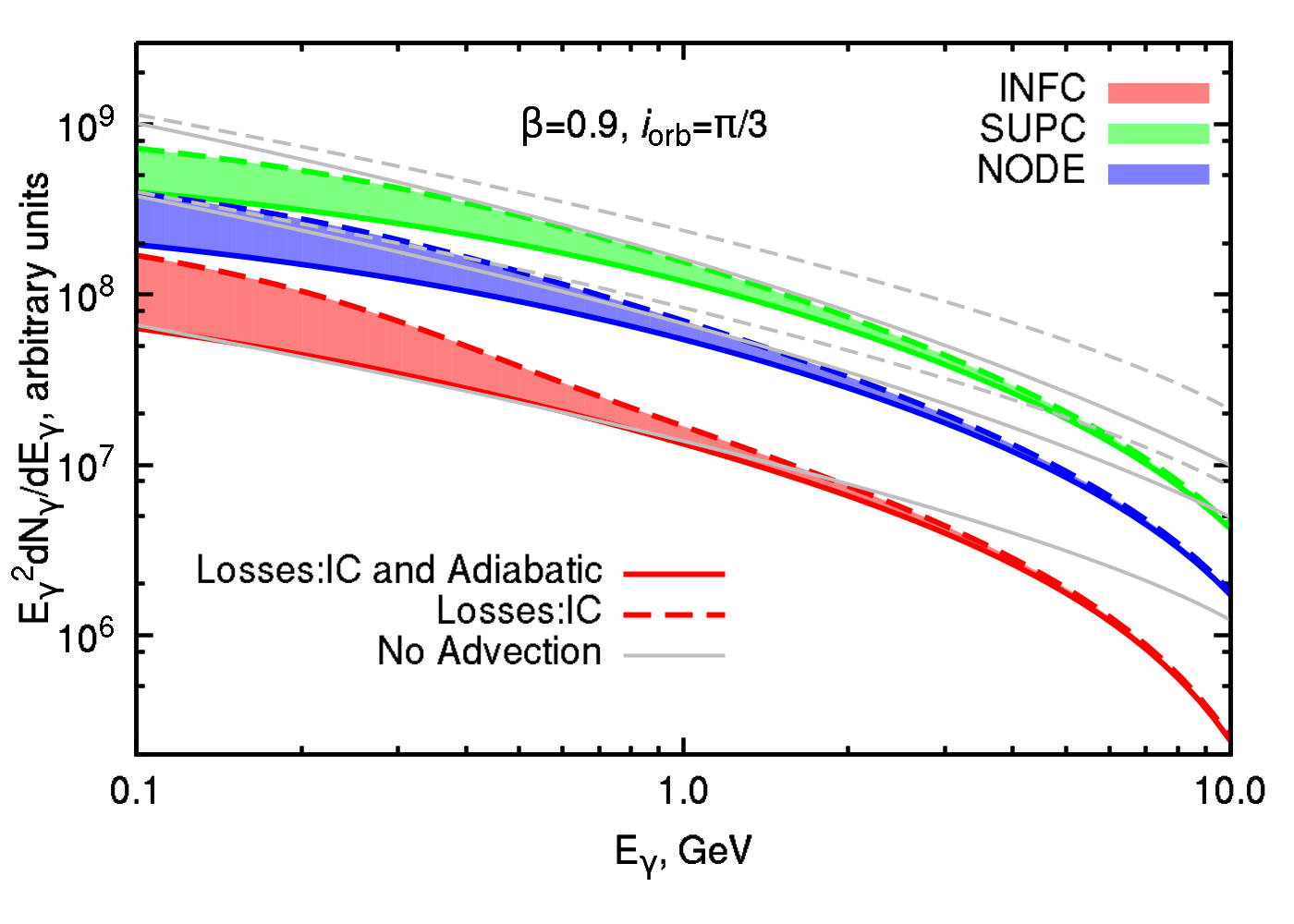}
  \caption{Spectral energy distributions of the IC emission for an extended emitter  obtained for three different orbital phases: SUPC (green); INFC (red); and NODE (blue). The jet is assumed to be perpendicular to the orbital plane, \(\alpha=\pi/2\). The case with weak adiabatic losses is shown with dashed lines, and adiabatic losses for a conical jet with solid lines. The region between these two regimes is filled with  color. Gray lines show the spectral energy distributions obtained for a compact emitter (Eq.~\ref{eq:spectrum_compact}). Gray dashed lines show the spectral energy distributions for SUPC and NODE phases obtained by applying the orbital-phase dependent coefficient in Eq.~\eqref{eq:spectrum_compact_v3} to INFC spectrum obtained under CE approximation (gray solid line). Cases with \(\beta=0.9\) and \(i_\textsc{orb}=30^\circ\), \(\beta=0.7\) and \(i_\textsc{orb}=30^\circ\), and \(\beta=0.9\) and \(i_\textsc{orb}=60^\circ\), are shown in the top, middle and bottom panels, respectively. Other model parameters were set as \(B=0\) and \(x_0=3d\).}
  \label{fig:ic}
\end{figure}

The combined effects of particle advection and changes in the photon density and scattering angle may affect the gamma-ray spectrum in a quite complex manner, as shown in Fig.~\ref{fig:ic}. For certain orientations of the jet, the emission from the inferior conjunction is strongly suppressed by unfavorable scattering angles, in which case advection along the jet will tend to enhance the emission below \(1\rm\,GeV\). {At high electron energies, the cooling length is shorter, and the IC cross-section has a weaker dependence on the scattering angle, so the particle transport  has a minor impact on the high-energy part of the spectrum. Thus, advection may distort the spectral the spectral shape in the GeV energy band. This effect is strongly pronounced in fast jets (see top and bottom panels of Fig.~\ref{fig:ic}). }

\begin{figure}
  \includegraphics[width=\columnwidth]{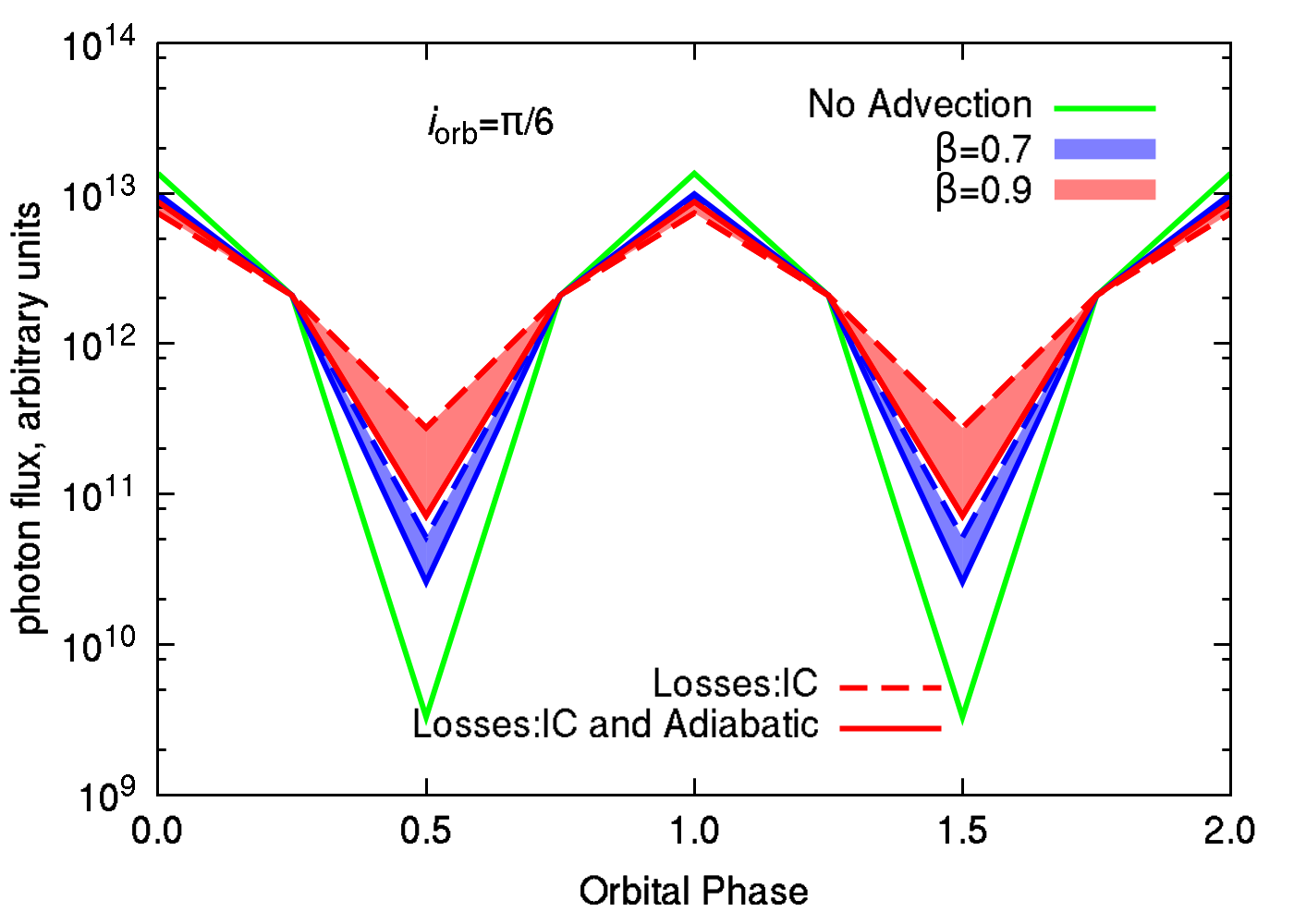}
  \includegraphics[width=\columnwidth]{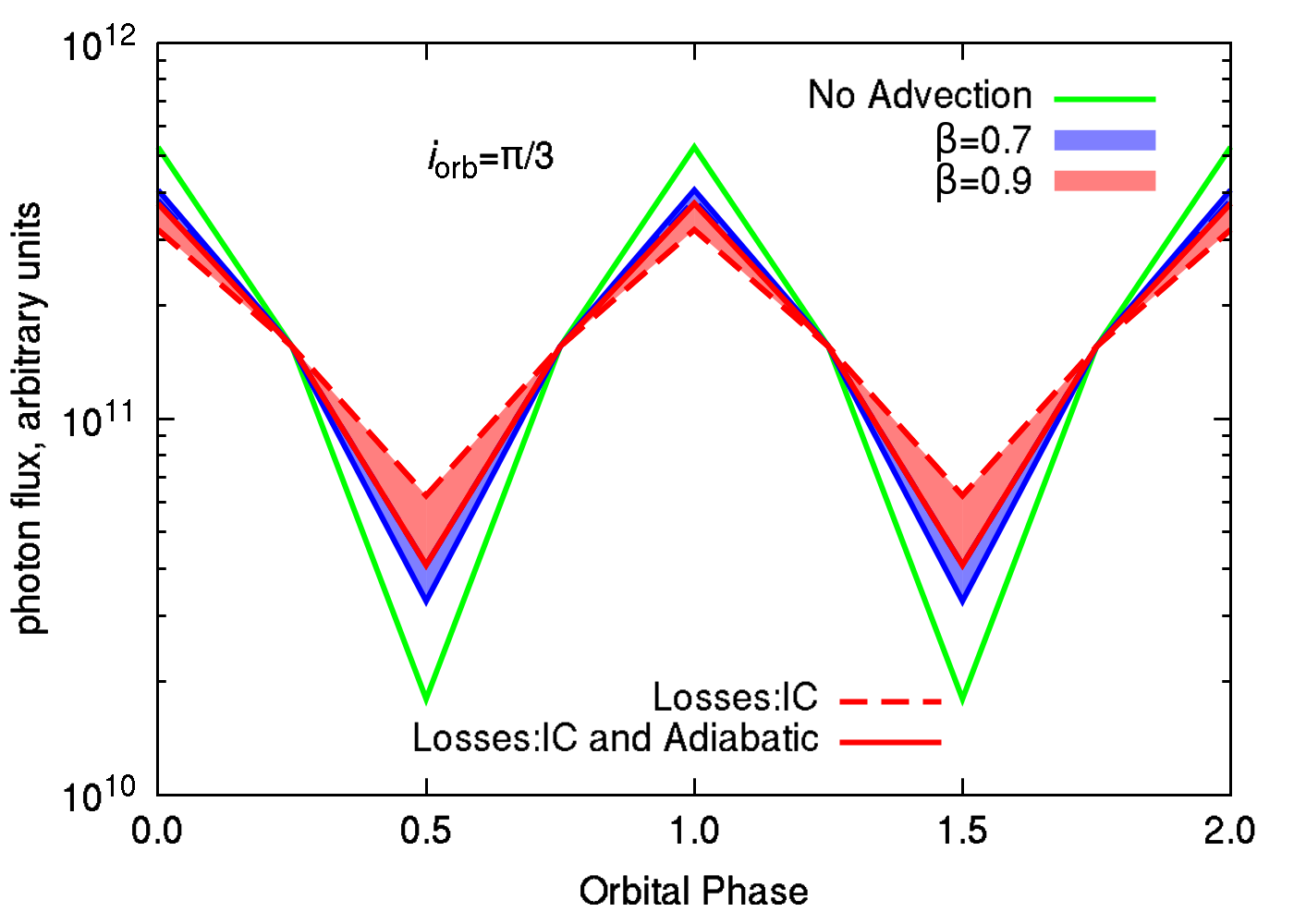}
  \caption{Lightcurves of the IC emission accounting for electron advection. Calculations are performed for two orbital inclinations: \(i_\textsc{orb}=60^\circ\) (top panel) and \(i_\textsc{orb}=30^\circ\) (bottom panel). The jet was assumed to be perpendicular to the orbital plane, \(\alpha=\pi/2\). The case with weak adiabatic losses is shown with dashed lines, and adiabatic losses for a conical jet with solid lines. The region between these two regimes is filled with  color. Two different jet velocities are shown with different colors: \(\beta=0.7\) (blue); and \(\beta=0.9\) (red). Green lines correspond to the electron spectrum obtained for a compact emitter(Eq.~\ref{eq:n_eff}). Other model parameters were set as \(B=0\) and \(x_0=3d\).}
  \label{fig:lc}
\end{figure}

Around the superior conjunction of the CO, advection tends to harden the  radiation spectrum. This is caused by the escape of lower energy electrons to regions of weaker photon fields, as illustrated in Fig.~\ref{fig:density_CygX3} with thin lines. As shown in Fig.~\ref{fig:ic}, advection tends to enhance the emission around inferior conjunction, and to reduce the emission around superior conjunction. Thus, the orbital variation of the gamma-ray emission will be weakened by advection. This is illustrated in Fig.~\ref{fig:lc}, where lightcurves for different jet velocities (\(\beta=0.7\) and \(\beta=0.9\)), and inclinations (\(i_\textsc{orb}=30^\circ\) and \(i_\textsc{orb}=60^\circ\)), are shown, with the acceleration site assumed to be at \(x_0=3d\). It can be seen that even in the slower case with \(\beta=0.7\), the orbital variation becomes significantly weaker. 

As expected, adiabatic losses result in a less extended production site, and thus the orbital phase dependence remains stronger than in the case of negligible adiabatic losses (solid vs dashed lines in Fig.~\ref{fig:lc}). However, the spectral change caused by propagation effects may remain strong. In particular, the combined effects of particle advection towards regions with weaker photon field, and adiabatic cooling, result in a hardening of the photon spectrum around \(0.1\)~-~\(1\)~GeV (see Fig.~\ref{fig:ic}).


\section{Quantum Electrodynamics Effects}\label{sec:qed}

There are two quantum electrodynamics (QED) effects that may have a substantial influence on the gamma-ray emission produced in compact binary systems. The first is related to the transition from the classical Thomson limit to the quantum Klein-Nishina regime. The Thomson limit is valid when electron and target photon energies are small:
\be
4 \ve \omega_{\rm ph}(1-\cos\theta_\textsc{sc})\ll1\,.
\ee
If the electron and the target photon energy are high enough to violate this relation, the precise QED cross-section should be used \citep[for astrophysical conditions, see][]{1981Ap&SS..79..321A}. The Klein-Nishina effect has a strong impact both on the energy loss rate and on the IC spectrum. 

\begin{figure}
  \includegraphics[width=\columnwidth]{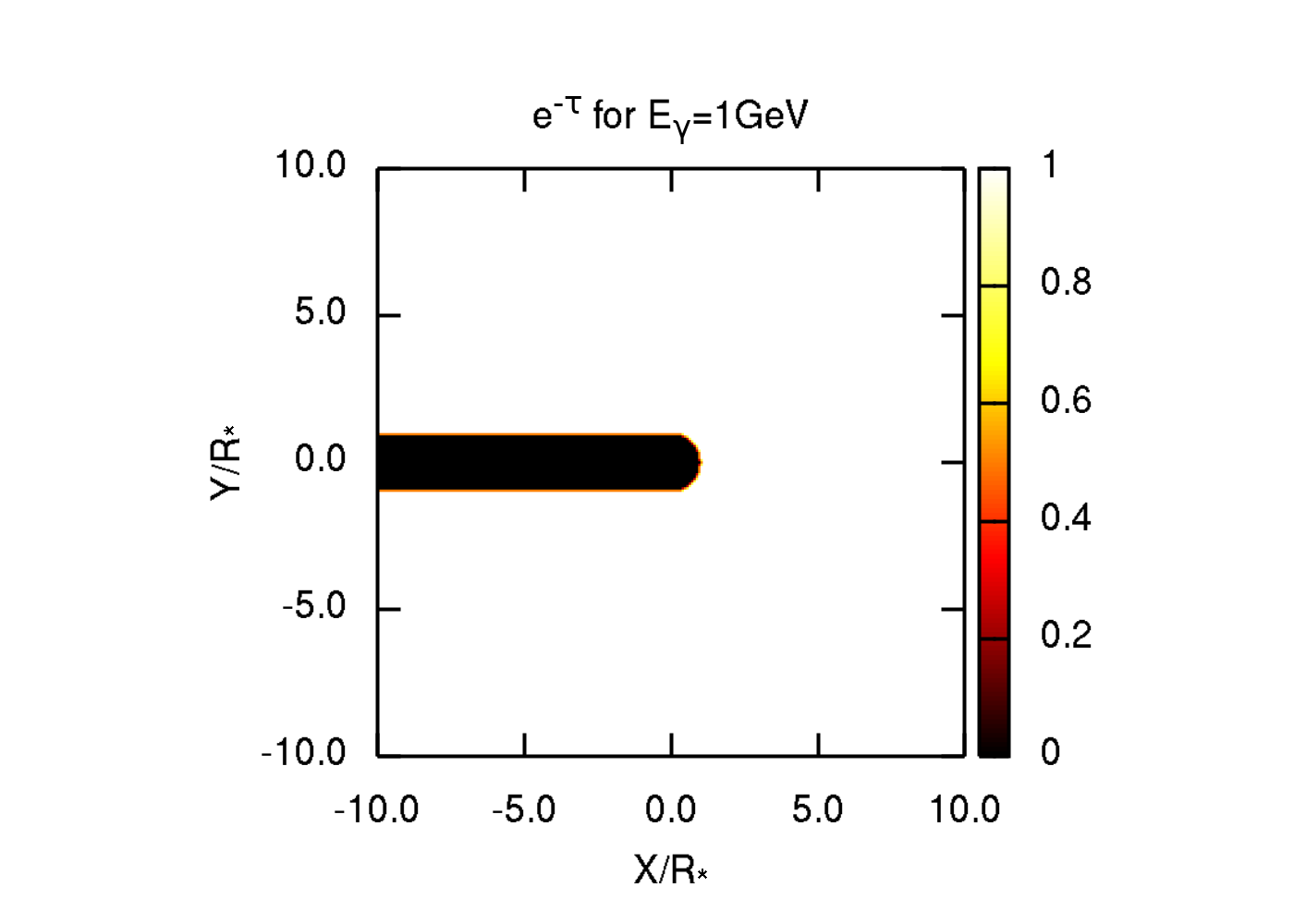}
  \includegraphics[width=\columnwidth]{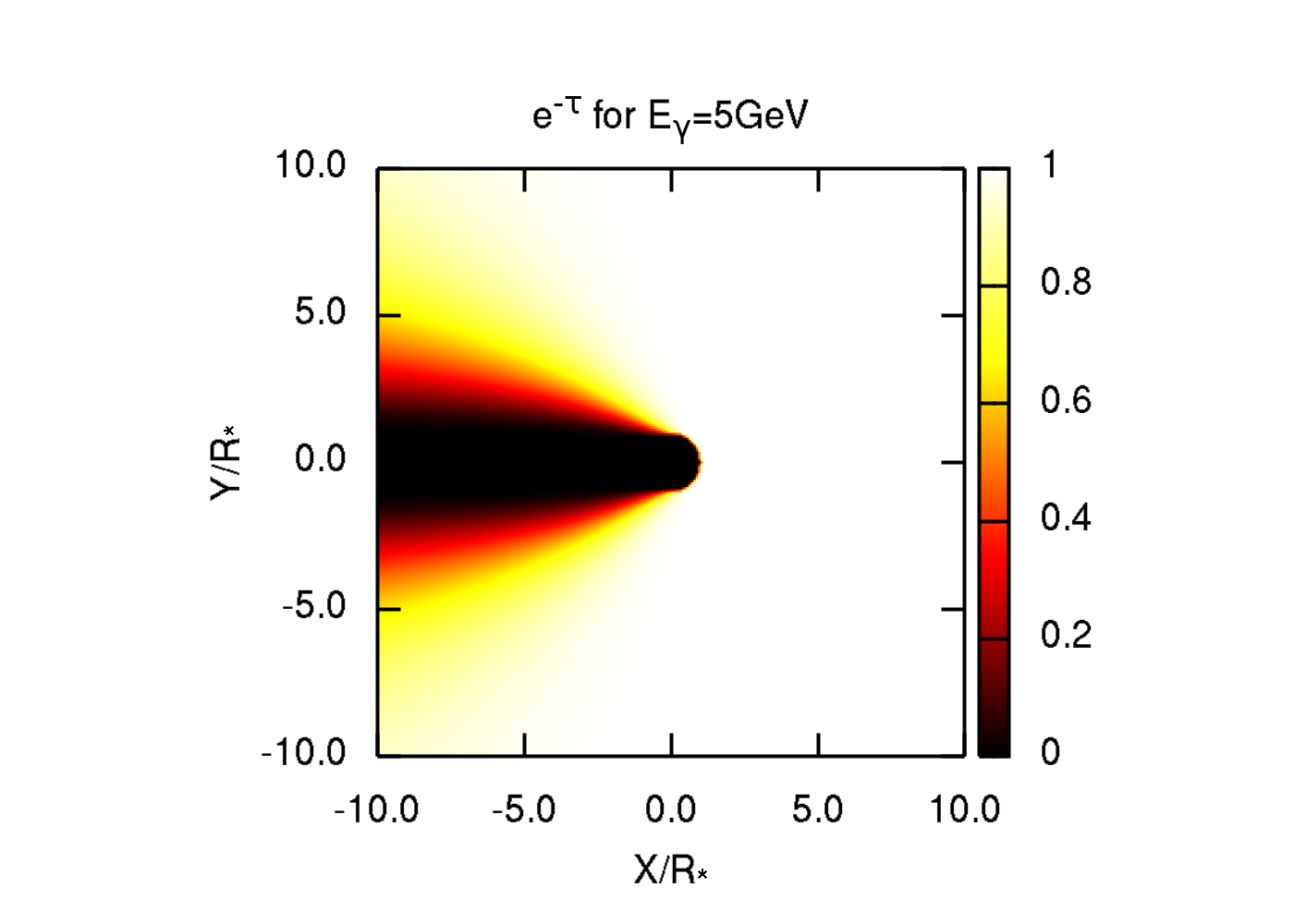}
  \includegraphics[width=\columnwidth]{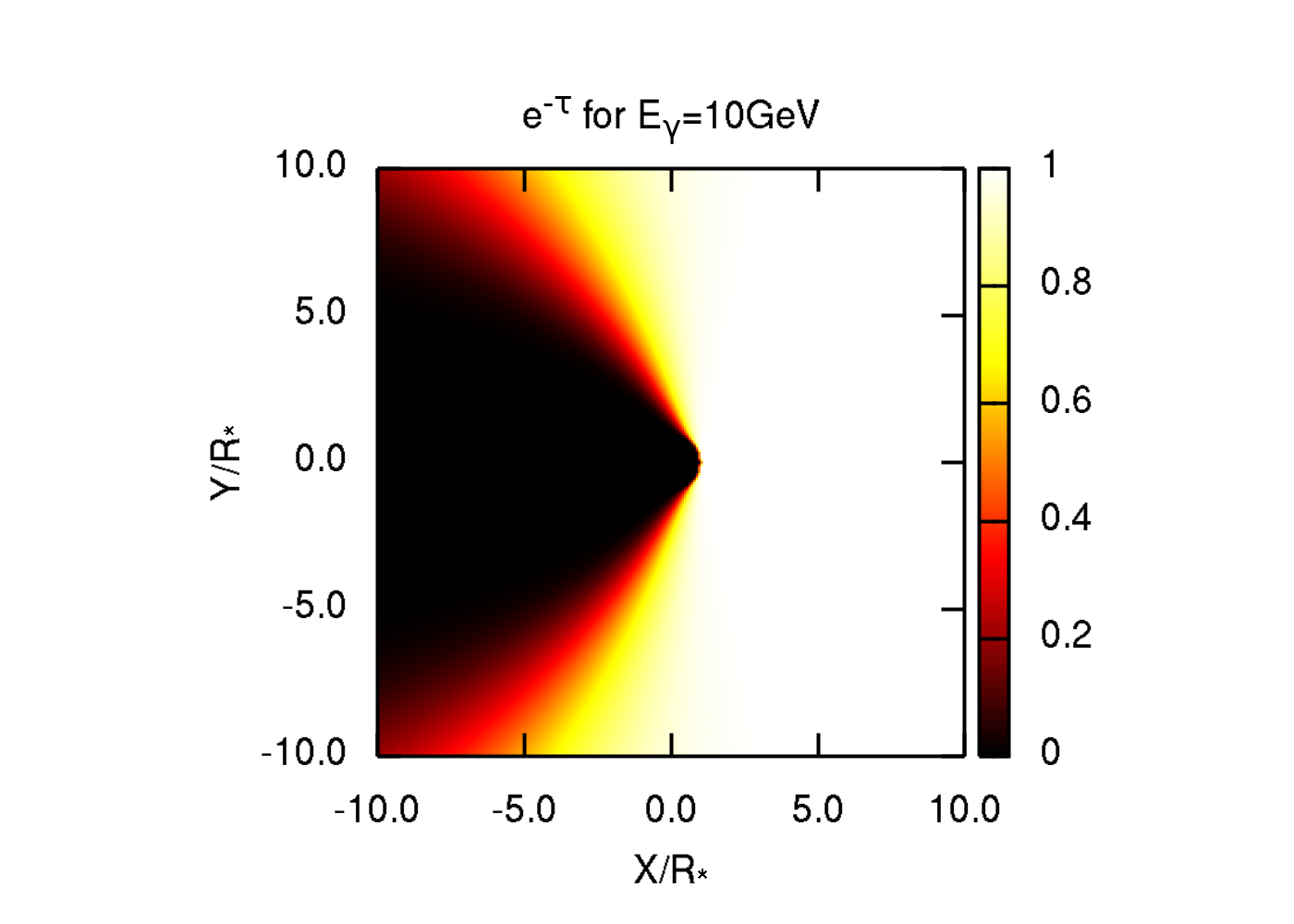}
  \caption{Gamma-gamma attenuation factor for 1~GeV (top panel), 5~GeV (middle panel), and 10~GeV (bottom panel) gamma rays traveling towards an observer looking from the right. The calculations are done for \(R_*=1.6\times10^{11}\rm\,cm\) and \(T_*=10^5\rm\,K\).}
  \label{fig:tau}
\end{figure}

The second important QED effect is the gamma-gamma attenuation. Typically, in binary systems this effect is important in the TeV energy band \citep{2006A&A...451....9D}. If the stellar temperature is high, as, e.g., in \cyg, the attenuation might be important for gamma rays with relatively low energy, \(E_\gamma\geq10\rm\,GeV\) \citep{1987ApJ...322..838P,1993MNRAS.260..681M,1997A&A...322..523B,2011A&A...529A.120C,2012MNRAS.421..512S}. In Fig.~\ref{fig:tau} we show the attenuation factor for gamma rays interacting with the stellar field. The target photon field is provided by the optical star with radius and temperature of  \(R_*=1.6\times10^{11}\rm\,cm\) and  \(10^5\rm \,K\), respectively. The calculation takes into account the finite size of the star integrating over the stellar surface  
\citep[which can give a substantial difference as compared to calculations adopting a point-like approximation for the star if the gamma-ray emitter locates within a few stellar radius distance from the star, see, e.g.,][]
{2006A&A...451....9D,2008A&A...482..397B,2009IJMPD..18..347B,2010A&A...518A..12R}; 
this accounts for the occultation by the star, as seen in the map opacity for \(1\rm\,GeV\) photons in Fig.~\ref{fig:tau}. For \(10\rm\,GeV\) gamma rays the attenuation can be very significant for almost half of the orbit unless the gamma-ray production site is located at a large distance from the CO. At a few GeV the influence of the gamma-ray absorption is smaller, although it still can suppress the emission from the counter-jet at SUPC phases, which can be relevant if the jet bulk velocity is relatively small. This may result in an additional factor affecting the orbital variability with a strong energy dependence, yielding a multi-GeV lightcurve significantly different from the GeV lightcurve.

\begin{figure}
  \includegraphics[width=\columnwidth]{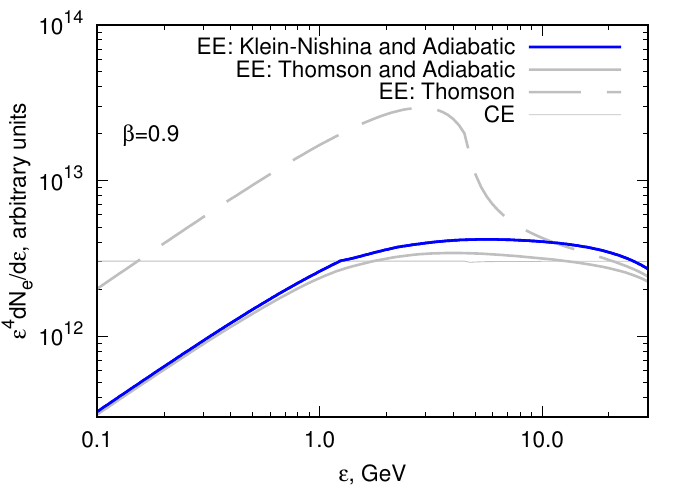}
  \caption{Energy distribution of electrons in the jet calculated for conditions similar to those in \cyg. The jet velocity was assumed to be \(\beta=0.9\) and nonthermal electrons with \(\alpha_\textsc{inj}=3\) were injected at a distance \(x=3d\) from the CO. Gray lines correspond to densities shown in Fig.~4 and obtained for Thomson losses: compact emitter (CE, thin line); extended emitter with Thomson losses only (dashed line); and extended emitter with adiabatic losses for a conical jet (solid line). The solid blue line shows the energy distribution of electrons computed using an accurate IC loss prescription, under adiabatic losses in a conical jet. The jet was taken perpendicular to the orbital plane, \(\alpha=\pi/2\).}
  \label{fig:density_kn}
\end{figure}

In the case of Klein-Nishina losses, electrons may lose a significant fraction of energy in a single interaction, which is inconsistent with the assumptions used for the continuous-loss approximation. However, the continuous-loss approximation was shown to provide results consistent with a detailed kinetic treatment \citep[see, e.g.,][]{2005AIPC..745..359K}. Thus, to account for the Klein-Nishina effect we solve Eq.~\eqref{eq:losses} for a conical jet using the approximation for IC losses in a Planckian photon fiedd suggested in \citet{2014ApJ...783..100K}. To compute the electron density, we use the continuous-loss approximation, i.e., the energy distribution density of electrons is described by Eq.~\eqref{eq:effective_density_gen}. The influence of the reduction of energy losses due to the Klein-Nishina effect is shown in Fig.~\ref{fig:density_kn}. It is seen that the weakening of the IC energy losses results in a \(\sim30\%\) increase of the number density of GeV electrons. 

\begin{figure}
  \includegraphics[width=\columnwidth]{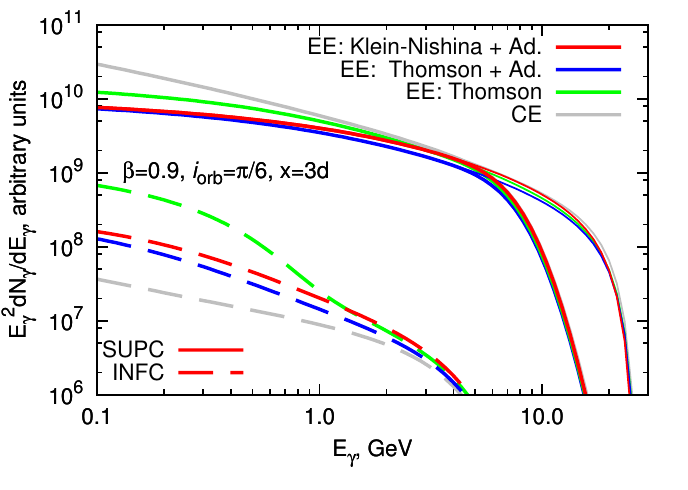}
  \includegraphics[width=\columnwidth]{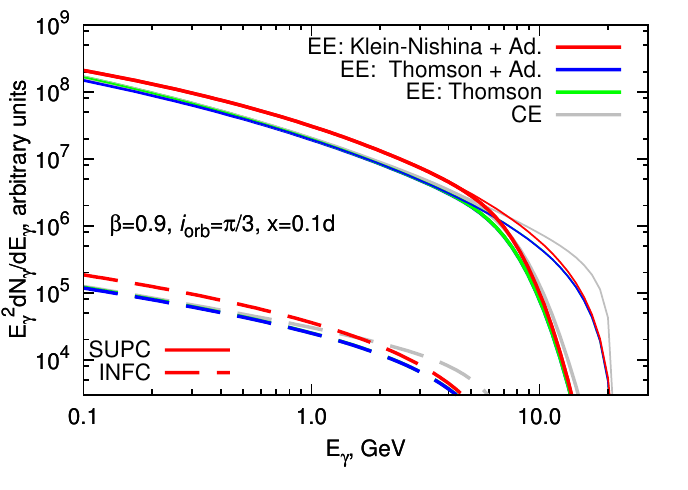}
  \caption{Comparison of gamma-ray spectral energy distributions calculated under different assumptions on transport and cooling for: emission obtained for a compact emitter (gray lines); dominant Thomson cooling (green lines); Thomson cooling with adiabatic losses in a conical jet (blue); and Klein-Nishina cooling with adiabatic losses in a conical jet (red lines). Two orbital phases are shown: SUPC (solid lines) and INFC (dashed lines). Calculations account for gamma-gamma attenuation in the photon field of the optical companion. For SUPC, the intrinsic spectra are shown with thin lines (for INFC the attenuation is negligible).  The jet velocity was assumed to be \(\beta=0.9\), and two different injection points are considered: \(x=3d\) (\(i_\textsc{orb}=\pi/6\), top panel) and \(x=0.1d\) ( \(i_\textsc{orb}=\pi/3\), bottom panel). The jet was taken perpendicular to the orbital plane, \(\alpha=\pi/2\).}
  \label{fig:ic_KN}
\end{figure}

For each considered gamma-ray energy and location in the jet we compute the gamma-gamma opacity, \(\tau_{\gamma\gamma}\), in the stellar photon field in the direction of the observer. The influence of different transport and cooling assumptions is illustrated in Fig.\ref{fig:ic_KN}. As seen in the figure, Klein-Nishina IC cooling has a similar impact at different orbital phases on the gamma-ray emission intensity, resulting in a small transformation of the lightcurve shape. In contrast, gamma-gamma attenuation strongly affects the gamma-ray spectrum above \(5\rm\,GeV\) for SUPC. We note that Fig.~\ref{fig:ic_KN} shows the emission produced in the jet; for the counter-jet the impact should be considerably stronger, for relatively low jet velocities. Thus, a detailed study of multi-GeV gamma-ray emission from \cyg may significantly constrain the possible locations of the production site. 

\section{Conclusions}

We have studied the properties of the gamma-ray emitting region in a relativisitc jet in a binary system. To facilitate the interpretation of the results, we have used an approach based on the distribution function in the phase space, which is Lorentz invariant. This allows obtaining results in a compact form that permits studying the influence of different parameters in a clearer way. The main focus of the study was on the impact of advection on the gamma-ray spectrum and lightcurve.

For the case of a compact production site we have obtained an analytic representation of the energy distribution of the emitting electrons. When IC cooling dominates over advection, the gamma-ray spectrum, given by Eq.~(\ref{eq:spectrum_compact}), has a
simple form that allows one to determine the process that affects the variability of the emission. Namely, it contains three factors that change with orbital phase: (i) IC proceeds in the anisotropic regime, and the scattering angle varies along the orbit \citep{2008MNRAS.383..467K,2010MNRAS.404L..55D,2012MNRAS.421.2956Z}; (ii) the Doppler boosting factor, \(\left[{\Db^{2\alpha_\gamma+1}\Gamma^{-1} }\right]\), which accounts for the relativistic transformation of radiation produced in a stationary jet \citep{1997ApJ...484..108S}; and (iii) in the case of dominant IC losses, an additional factor, \(\Db_*^2\), should be introduced. The stellar photon boosting effect on cooling can be ignored if the dominant losses are due to synchrotron cooling.

{Adiabatic losses can be relevant only if relativistic particles are advected along the jet over a distance in which the jet material density undergoes a significant change.} In particular, this can be the case for low-energy electrons that are subject to slower radiative losses. In the case of a (at least) mildly relativistic jet, \(\Gamma\geq2\), advection might be important for GeV emitting electrons even in the most compact binaries like \cyg. In the case of dominant radiative losses, we have obtained an analytic solution that describes the properties of non-thermal electrons in a relativistic inclined jet. This solution can however be generalized to the case when adiabatic losses are important under weak IC losses, i.e., covering a broad range of synchrotron and adiabatic losses. 

It is generally expected that in gamma-ray emitting \muq IC losses should dominate over synchrotron for GeV electrons \citep[see, e.g.,][]{2012MNRAS.421.2956Z}. Thus, as test cases, we have considered two cases for extended emitters: (i) dominant IC losses, which allow an analytic solution for the particle density, Eq.~\eqref{eq:extended_solution}; and (ii) the case with IC and adiabatic losses, the latter being expected in a conical jet, for which a numerical treatment has been applied. The simulations have shown that, in systems similar to \cyg, particle advection may have a significant impact on the gamma-ray lightcurve if the jet velocity is high, \(\beta\geq 0.7\). For even faster jet velocities, \(\beta\sim0.9\), one should also expect a strong transformation of the gamma-ray spectrum from different orbital phases. In a more extended system, e.g., in \cygl, advection is very important unless  synchrotron losses prevent efficient particle transport (see Eq.~\eqref{eq:b_dominant}), which is probably not very realistic.

In the specific case of \cyg, the stellar companion should be very hot, \(T_*\simeq10^5\rm\,K\). For such a target photon field, two QED effects may influence the electron transport and gamma-ray spectrum in the GeV energy band. The Klein-Nishana effect weakens the IC energy losses and affect the gamma-ray spectrum, and gamma-gamma absorption can significantly suppress the flux above a few GeV. To study the influence of these effects we have performed detailed calculations of the electron transport, radiation, and gamma-gamma opacity. Since in the case of \cyg, the orbital separation is comparable to the stellar radius, in the calculations of the gamma-gamma opacity we have accounted for the finite size of the optical star. The simulations show that the Klein-Nishina effect has a small impact on the intrinsic gamma-ray spectra. Unless the gamma-ray production site is located at large distance from the CO, \(x\gg3d\), the gamma-gamma attenuation should significantly affect the spectrum at multi-GeV energies, \(E_\gamma>5\rm\,GeV\). 

To summarize, we have performed a detailed study of the IC process in realistic jets in compact binary systems. The
performed study has revealed that the particle advection along the jet might be important even in a very compact binary
system, e.g., in \cyg. In systems similar to \cygl, advection should be accounted for even in the case of a weakly
relativistic jet. 

If adiabatic losses are weak, which would be the case, e.g., in cylindrical jets, advection can impact significantly the gamma-ray emission, potentially leading to a strong dependence of the gamma-ray spectrum shape on the orbital phase. For advection in a conical jet, adiabatic losses weaken the effects on the spectrum. 

Independently of the dominant cooling channel, advection results in a significant weakening of the orbital phase dependence. Thus, if the properties of the accelerator in \cyg and \cygl are similar,  one should expect differences in the orbital phase dependency of the GeV emission between these two systems.  

\begin{figure}
  \includegraphics[width=\columnwidth]{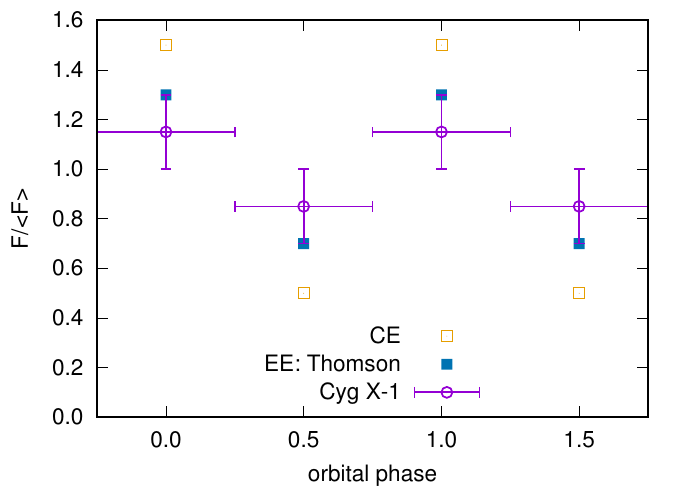}
  \caption{Lightcurves of the IC emission from a system similar to \cygl: \(T_*=3\times 10^4\rm\,K\), \(L_*=8\times10^{38}\,\ergs\),\(d=3.2\times10^{12}\rm \,cm\) (circular orbit), \(\beta=0.5\), and \(i_\textsc{orb}=60^\circ\). The jet was assumed to be perpendicular to the orbital plane, \(\alpha=\pi/2\). The case with weak adiabatic losses is shown with filled squares and the emission expected from a CE is shown with open squares. The data points are adopted from \citet{2017MNRAS.471.3657Z}. Other model parameters were set as \(B=0\) and \(x_0=4d\).}
  \label{fig:cygx1_lc}
\end{figure}

{To illustrate the relevance of this effect, in Fig.~\ref{fig:cygx1_lc} we show the lightcurves computed for a
  system similar to \cygl (the temperature and luminosity of the optical star are taken as \(T_*=3\times 10^4\rm\,K\)
  and \(L_*=8\times10^{38}\,\ergs\), respectively; the CO was assumed to be in a circular orbit with
  \(d=3.2\times10^{12}\rm \,cm\)).

  The IC emission shown in Fig.~\ref{fig:cygx1_lc} was averaged over two orbital phase bins: \(|\phi|<0.25\) and
  \(|\phi|>0.25\), the orbit being \(-0.25<\phi<0.75\)). The injection point was assumed to be located at \(x_0=4d\),
  and the injection spectrum and jet velocity were assumed to be \(\propto\ve^{-4}\) and \(\beta=0.5\),
  respectively. The orbital inclination was selected to be \(i_\textsc{orb}=\pi/3\) and the figure includes only the
  contribution from the jet (i.e., the counter-jet emission is not accounted for because it is expected to be relatively
  small and to weaken the orbital phase dependence even stronger). The data points are from \citet{2017MNRAS.471.3657Z},
  and the open and filled squares correspond to the emission expected from a CE and a EE, respectively. The adiabatic
  losses were assumed to be weak and the magnetic field set to \(B=0\), so the dominant cooling mechanism is the Thomson
  scattering.  As seen from Fig.~\ref{fig:cygx1_lc}, the advection may provide a possible explanation for a weaker
  orbital phase dependence of the GeV emission from \cygl, and alleviate the requirement for a SSC contribution.
  \citet{2017MNRAS.471.3657Z} studied the broadband emission and gamma-ray variability in \cygl for the parameter space
  with \(x_0\ll d\). In that parameter space, it was found that external Compton models, including those with an
  extended emitter, are incompatible with the \fer emission from \cygl, and a better agreement can be achieved if one
  assumes a highly clumpy jet, which enhances the SSC emission.

  In the case of a conical jet \citep[as was assumed by][]{2017MNRAS.471.3657Z}, adiabatic losses lead to considerable
  cooling at distances \(x\sim x_0\), so plasma cools down on a scale in which the IC regime does not change for
  \(x_0\ll d\). Thus, advection cannot considerably affect the IC lightcurve for parameters adopted by
  \citet{2017MNRAS.471.3657Z}.  The simulations shown in Fig.~\ref{fig:cygx1_lc} show that for \(x_0\geq d\), advection
  can improve the agreement between the observational data from \cygl and predictions of models that account for stellar
  IC only. We note, however, that the calculations presented in Fig.~\ref{fig:cygx1_lc} are for illustrative
  purposes only and cannot substitute a detailed broadband study (as the one presented in \citealt{2017MNRAS.471.3657Z}).
}

The influence of advection on the gamma-ray light curve also significantly affects the ability of one-zone models \citep{2010MNRAS.404L..55D,2012MNRAS.421.2956Z,2018MNRAS.479.4399Z} to accurately infer the properties of the gamma-ray production sites even in the case of the most compact binary systems. For example,  \citet{2018MNRAS.479.4399Z} suggested that the \fer emission in \cyg is best explained by IC scattering from a production site located at \(x=2.3d\) in a jet with \(\beta=0.73\). As shown by our simulations, for this location of the production site the transport effects might be relevant. 

We present in this paper the theoretical framework and discuss the impact of advection on the GeV gamma-ray spectrum and lightcurve. A detailed application to the gamma-ray data of \cyg will be presented in a forthcoming paper.

\section*{Acknowledgements}
We want to thank the anonymous referee for useful and constructive comments and suggestions.
This work was supported by JSPS KAKENHI Grant Number JP18H03722. 
V.B-R. acknowledges support by the Spanish Ministerio de Econom\'{i}a y Competitividad (MINECO/FEDER, UE) under grants AYA2016-76012-C3-1-P, with partial support by the European Regional Development Fund (ERDF/FEDER), and MDM-2014-0369 of ICCUB (Unidad de Excelencia `Mar\'{i}a de Maeztu'), and the Catalan DEC grant 2014 SGR 86.


\appendix
\bsp	
\label{lastpage}
\end{document}